\documentclass[%
11pt,
reprint,
onecolumn,
tightenlines,
superscriptaddress,
preprintnumbers,
nofootinbib,
amsmath,amssymb,amsthm,
physrev,
eqsecnum,tikz,
]{revtex4-2}

\usepackage{isomath}
\usepackage{amsmath,amsthm}
\usepackage{amsbsy}
\usepackage{amssymb}
\usepackage{amscd}
\usepackage{amsfonts}
\usepackage{stmaryrd}
\usepackage{siunitx}
\usepackage{euscript}
\usepackage[utf8]{inputenc}
\usepackage[T1]{fontenc}
\usepackage{newtxtext} 
\everymath{\displaystyle}
\usepackage{exscale}

\usepackage{graphicx}
\usepackage{boxedminipage}
\usepackage{calc}
\usepackage[usenames,dvipsnames]{xcolor}
\usepackage[caption=false,justification=centerlast]{subfig}

\usepackage{setspace}
\usepackage{enumitem}
\setitemize{noitemsep,topsep=0pt,parsep=0pt,partopsep=0pt}
\setenumerate{noitemsep,topsep=0pt,parsep=0pt,partopsep=0pt}
\setdescription{noitemsep,topsep=0pt,parsep=0pt,partopsep=0pt}

\usepackage{hyperref}
\usepackage[normalem]{ulem}

\usepackage[small]{titlesec}

\titlespacing*{\section}{0pt}{12pt plus 4pt minus 2pt}{2pt plus 2pt minus 2pt}
\titlespacing*{\subsection}{0pt}{12pt plus 4pt minus 2pt}{2pt plus 2pt minus 2pt}
\titlespacing*\subsubsection{0pt}{12pt plus 4pt minus 2pt}{2pt plus 2pt minus 2pt}
\titlespacing*\paragraph{0pt}{12pt plus 4pt minus 2pt}{2pt plus 2pt minus 2pt}

\makeatletter

    \renewcommand*{\p@subsection}{}
    
    \renewcommand*{\p@subsubsection}{}
\makeatother

\usepackage{isomath}
\usepackage{amsmath}
\usepackage{amssymb}
\usepackage{amscd}
\usepackage{amsfonts}

\newcommand{\half}{\frac{1}{2}}

\DeclareMathOperator{\trace}{tr}

\newcommand{\dm}{\ \mathrm{d}}
\newcommand{\Dm}{\ \mathrm{D}}

\newcommand{\bfr}{{\mathbold r}}

\newcommand{\bfx}{{\mathbold x}}

\newcommand{\bfF}{{\mathbold F}}

\newcommand{\bfI}{{\mathbold I}}

\newcommand{\bfP}{{\mathbold P}}

\newcommand{\bfR}{{\mathbold R}}

\newcommand{\bfU}{{\mathbold U}}

\newcommand{\bfX}{{\mathbold X}}

\usepackage{xcolor}

\usepackage{siunitx}

\begin{document}

\preprint{Phys. Rev. E 107, 064501 (DOI: 10.1103/PhysRevE.107.064501)}

\title{Statistical Field Theory for Nonlinear Elasticity of Polymer Networks with Excluded Volume Interactions}

\author{Pratik Khandagale}
    \email{pratik.khandagale701@gmail.com}
    \affiliation{Department of Mechanical Engineering, Carnegie Mellon University}

\author{Timothy Breitzman}
    \affiliation{Air Force Research Laboratory}

\author{Carmel Majidi}
    \affiliation{Department of Mechanical Engineering, Carnegie Mellon University}
    \affiliation{Department of Civil and Environmental Engineering, Carnegie Mellon University}
    \affiliation{Department of Materials Science and Engineering, Carnegie Mellon University}

\author{Kaushik Dayal}
    \affiliation{Department of Civil and Environmental Engineering, Carnegie Mellon University}
    \affiliation{Department of Mechanical Engineering, Carnegie Mellon University}
    \affiliation{Center for Nonlinear Analysis, Department of Mathematical Sciences, Carnegie Mellon University}
    \affiliation{Pittsburgh Quantum Institute, University of Pittsburgh}

\date{\today}


\begin{abstract}
    Polymer networks formed by cross-linking flexible polymer chains are ubiquitous in many natural and synthetic soft matter systems.
    Current micromechanics models generally do not account for excluded volume interactions except, for instance, through imposing a phenomenological incompressibility constraint at the continuum-scale.
    This work aims to examine the role of excluded volume interactions on the mechanical response.
    
    The approach is based on the framework of the self-consistent statistical field theory of polymers, which provides an efficient mesoscale approach that enables the accounting of excluded volume effects without the expense of large-scale molecular modeling.
    A mesoscale representative volume element is populated with multiple interacting chains, and the macroscale nonlinear elastic deformation is imposed by mapping the end-to-end vectors of the chains by this deformation.
    In the absence of excluded volume interactions, it recovers the closed-form results of the classical theory of rubber elasticity.
    With excluded volume interactions, the model is solved numerically in 3-dimensions using a finite element method to obtain the energy, stresses, and linearized moduli under imposed macroscale deformation.
    Highlights of the numerical study include: 
    (1) the linearized Poisson's ratio is very close to the incompressible limit without a phenomenological imposition of incompressibility;
    (2) despite the harmonic Gaussian chain as a starting point, there is an emergent strain-softening and strain-stiffening response that is characteristic of real polymer networks, driven by the interplay between the entropy and the excluded volume interactions;
    and (3) the emergence of a deformation-sensitive localization instability at large excluded volumes.
\end{abstract}

\maketitle


\section{Introduction}

    A wide variety of soft-matter based systems are emerging as important for engineering and scientific applications, and have been the focus of research using both modeling and experiments, e.g. \cite{deng2014electrets, chen2021interplay, ahmadpoor2015flexoelectricity, kim2016foam, brown2009multiscale, su2012semiflexible, grasinger2021architected, grasinger2021nonlinear, grasinger2020statistical, cohen2016electroelasticity, darbaniyan2019designing, grasinger2021flexoelectricity, grasinger2022group, markvicka2018autonomously, bartlett2017high, kazem2017soft, majidi2019soft, zhao2021modeling, bartlett2019self, ford2019multifunctional, zolfaghari2020network, ohm2021electrically, malakooti2020liquid, ware2016localized, white2015programmable, ware2015voxelated, ambulo2017four, mu2019sheath, saed2019molecularly, wie2016photomotility, babaei2017steering, babaei2021torque}.
    Polymer-network based materials such as elastomers and hydrogels are often at the heart of these soft-matter systems.

\begin{figure*}[ht!]
	\includegraphics[width=0.47\textwidth]{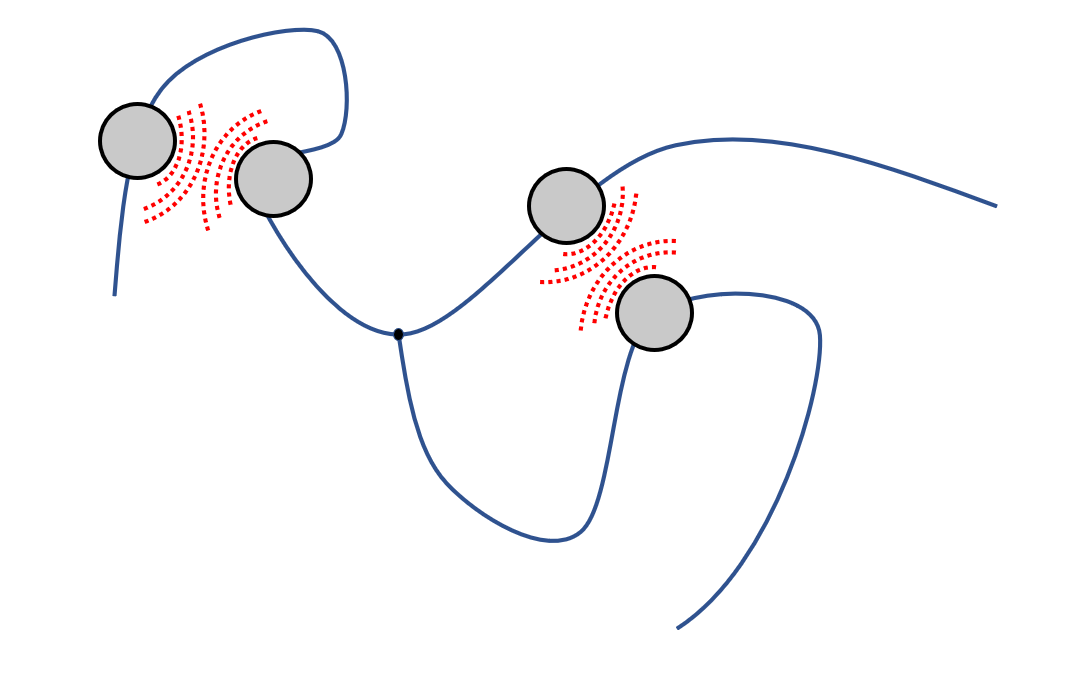}
	\caption{Excluded volume interactions are nonlocal in terms of the segment coordinates.}
	\label{fig:nonlocal-interaction}
\end{figure*}

An important question for both fundamental understanding and application is to predict the nonlinear elastic properties of polymer networks starting from a micromechanical model of individual chains.
The physics of polymer network elasticity is governed by the conformational entropy of polymer chains and the inter-segment excluded volume interactions. 
These contributions can be roughly thought of as short-range and nonlocal interactions respectively.
The short-range interactions are associated with Gaussian polymer chain response and depend on the relative configurations of adjacent segments in a chain.
In contrast, the nonlocal interactions are due to the interaction between polymer segments that are nearby in space but nonlocal topologically (i.e., in terms of their position along the chain (Fig. \ref{fig:nonlocal-interaction})).

While there are several useful phenomenological nonlinear elastic frame-indifferent models, e.g. Mooney-Rivlin \cite{mooney1940theory}, Ogden \cite{ogden1997non} and Gent \cite{gent1996new}, they lack a clear connection to the molecular structure of polymer network.
An important class of physics-based approaches to study the elasticity of polymer networks are based on considering multiple Gaussian chains and then averaging over the chains in different ways.
These include the 3-chain model by James and Guth \cite{james1943theory}, the 4-chain model by Flory and Rehner \cite{flory1943statistical} and Treloar \cite{treloar1946elasticity}, the affine full-network model by Treloar \cite{treloar1954photoelastic}, the 8-chain model by Arruda and Boyce \cite{arruda1993three}, the non-homogeneous deformation based model by Wu and van Der Giessen \cite{wu1993improved}, and the non-affine microsphere model by Miehe \cite{miehe2004micro}; the recent work by Grasinger \cite{grasinger2023networks} provides a new perspective in which these myriad models are shown to be special cases of a general approach.
While these models have provided important insights and prediction, they do not account for the nonlocal excluded volume effects.
Consequently, incompressibility of the polymer network must be added as a phenomenological continuum-scale approximation of the missing mesoscale physics.
Another class of physics-based molecular-statistical approaches are the constrained junction and constrained segment theories, that aim to account for constraints arising due to chain entanglements.
Constrained junction theories, e.g. \cite{ronca1975approach,flory1976statistical,flory1977theory,erman1978theory,flory1982theory}, apply topological constraints on the fluctuations of chain cross-link junctions. 
Constrained segment theories, e.g. \cite{deam1976theory,edwards1988tube,heinrich1983strength,heinrich1984strength,heinrich1988rubber}, which are consistent with the tube model of rubber elasticity, incorporate constraints on the polymer segments along the chain contour. 
However, it is not easy to incorporate the nonlocal excluded volume interactions in these approaches.

\subsection{The Proposed Approach}

Our approach is composed of 2 key elements: first, the statistical field theory of polymers which provides an established and efficient approach to account for the physics of polymer chain elasticity as well as excluded volume interactions \cite{de1969some,de1979scaling,matsen2006self,doi1986theory,fredrickson2006equilibrium,fredrickson2002field,chantawansri2007self,lennon2008free,drolet1999combinatorial,sides2006hybrid,ackerman2017finite,fredrickson2007computational, delaney2016recent, cochran2006stability, lennon2008numerical, matsen1994stable, matsen2001standard}; and, second, the use of the 8-chain network averaging model that provides a nonlinearly-elastic frame-indifferent approach to coarse-grain to the continuum scale \cite{arruda1993three}.
An important work in this direction is \cite{schmid2013self}, wherein a network with a simplified square lattice topology was studied using the field theory approach to understand copolymers.

\begin{figure*}[ht!]
	\subfloat[Schematic of a polymer network.]{\includegraphics[width=0.47\textwidth]{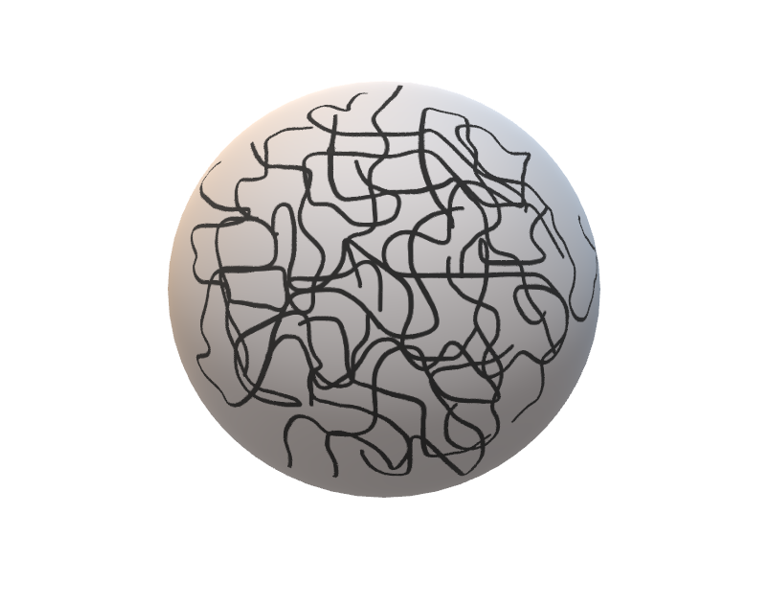}}
	\hfill
	\subfloat[The 8-chain approximation.]{\includegraphics[width=0.47\textwidth]{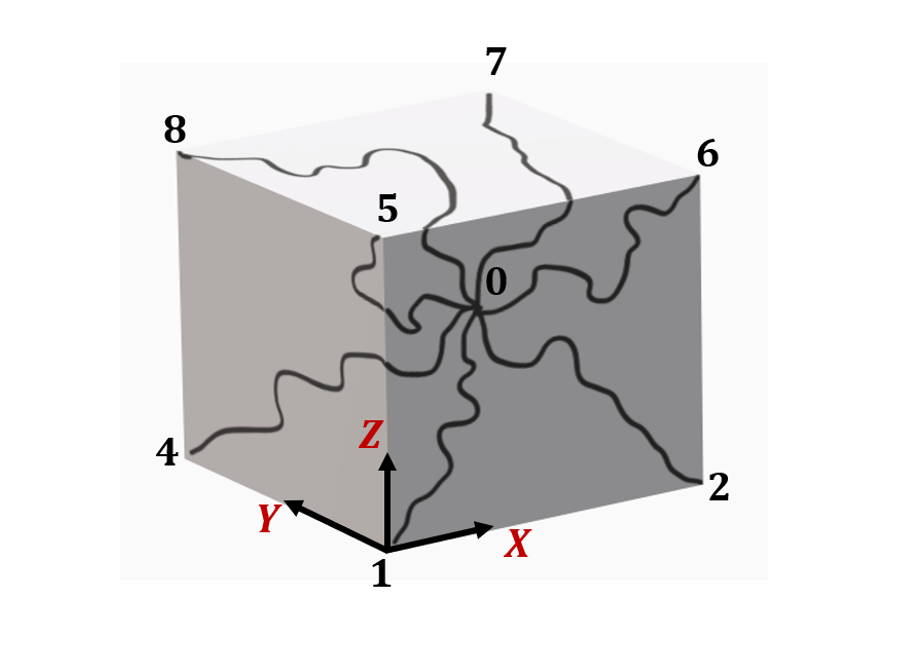}}
	\caption{The 8-chain approximation is obtained by averaging over a volume element that aligns the polymer chains along the principal directions of the deformation.}
	\label{fig:8-chain}
\end{figure*}

\begin{figure*}[ht!]
	\includegraphics[width=0.95\textwidth]{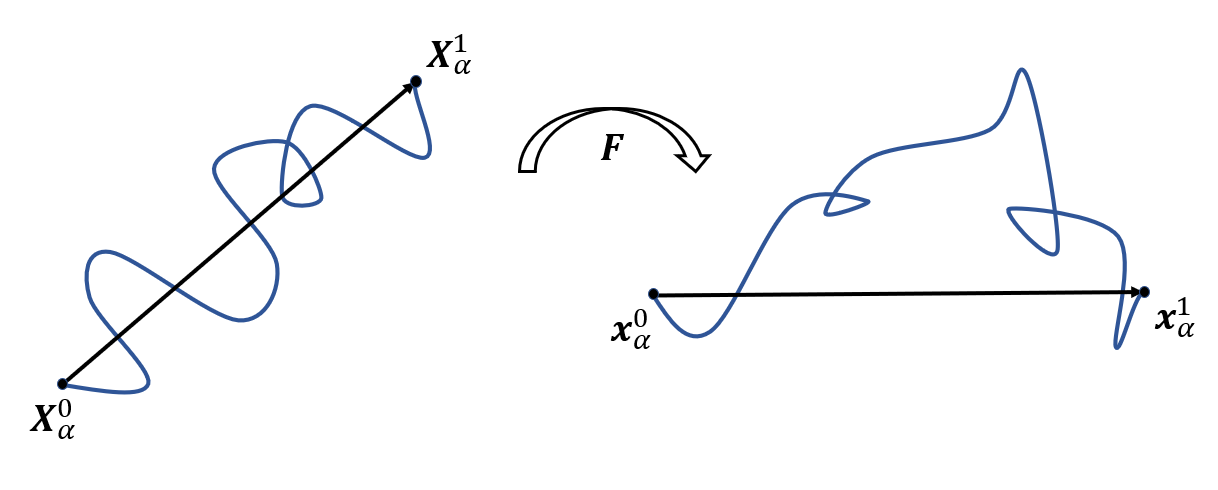}
	\caption{The end-to-end vector in the undeformed state, $\left(\bfX_{\alpha}^1- \bfX_{\alpha}^0\right)$, is mapped by the deformation $\bfF$ to the end-to-end vector in the deformed state, $\bfx_{\alpha}^1- \bfx_{\alpha}^0$, i.e., $\bfx_{\alpha}^1- \bfx_{\alpha}^0=  \bfF  \left(\bfX_{\alpha}^1- \bfX_{\alpha}^0\right)$.}
	\label{fig:affine-deformation}
\end{figure*}

We begin by considering a representative volume element (RVE) of the polymer network.
A typical mesoscale RVE consists of several polymer chains that are all interacting with configurations that are randomly distributed.
While it is a significant challenge to account for this randomness, we follow the 8-chain RVE-averaging approach of Arruda and Boyce (Fig. \ref{fig:8-chain}, \cite{arruda1993three}) in approximating the RVE in the undeformed state as composed of 8 polymer chains connecting the center
 of a cube to each of the corners.
The RVE then deforms under the action of the macroscopic deformation tensor $\bfF$, i.e., the chain end-to-end vectors are mapped by $\bfF$ from the undeformed to the deformed state (Fig. \ref{fig:affine-deformation}, \cite{treloar1975physics}).
An important element of \cite{arruda1993three} is that the RVE is oriented such that the cube is oriented along the principal directions of the stretch tensor $\bfU$, where $\bfU$ is the tensor square root of $\bfF$, and can be obtained through the polar decomposition $\bfF = \bfR \bfU$ where $\bfR \in SO(3)$.

Given the mapping of the end-to-end vectors of the chains, the polymer field theory is then used to compute the partition function of the deformed state, from which we can find the free energy and stress.
Following \cite{fredrickson2002field}, we use the continuous Gaussian chain model for a single polymer chain.
Next, we consider chain segments that interact pairwise in real-space -- and nonlocally in terms of position along the chain contour (Fig. \ref{fig:nonlocal-interaction}) -- through a pairwise interaction potential of mean force; these are given by Dirac potentials to model excluded volume effects.
Given the inter-segment interaction and end-to-end vectors, the framework of polymer field theory enables us to compute, using the self-consistent scheme, the partition function and consequently the free energy of the RVE.
We notice that because the ends of the polymer chain are constrained by the macroscale deformation $\bfF$, this leads to a restricted ensemble.
Further, nonlinear elasticity provides the Piola-Kirchoff stress tensor $\bfP$ as the energy-conjugate of $\bfF$, enabling us to compute the stress-deformation response of the polymer network.

Key results from the model are as follows.
In the absence of excluded volume interactions, we find that the closed-form orientationally-averaged elastic response matches with classical rubber elasticity \cite{treloar1975physics}.  
Considering excluded volume interactions, closed-form solutions appear impossible, and we develop a 3-d finite element method (FEM) implementation to self-consistently solve the equations of the polymer field theory.
We find that the linearized Poisson's ratio $\nu \simeq 0.4943$, which is very close to the incompressible limit $\nu\to 0.5$, without a phenomenological imposition of incompressibility, and that the elastic moduli are in line with typical polymer network gels.
Further, despite the harmonic Gaussian chain as a starting point, there is an emergent strain-softening and strain-stiffening response that is characteristic of real polymer networks, driven by the interplay between the excluded volume interactions and the entropy; it does not require chains with limiting extensibility -- such as the inverse Langevin approximation -- to model this behavior.
Finally, we find the emergence of a deformation-sensitive localization instability at large values of the excluded volume parameter.

\paragraph*{Structure of the Paper.}
Section \ref{sec:mathematical formulation} formulates the model.
Section \ref{sec:numerical method} summarizes the finite element approach for the self-consistent solution.
Section \ref{sec: results and discussion} presents numerical results showing the predictions of the model.


\section{Model Formulation }\label{sec:mathematical formulation}

\subsection{Deformation of a Single Polymer Chain}\label{sec:deformation of single polymer chain}

\begin{figure}[ht!]
	\centering
    \includegraphics[width=0.70\textwidth]{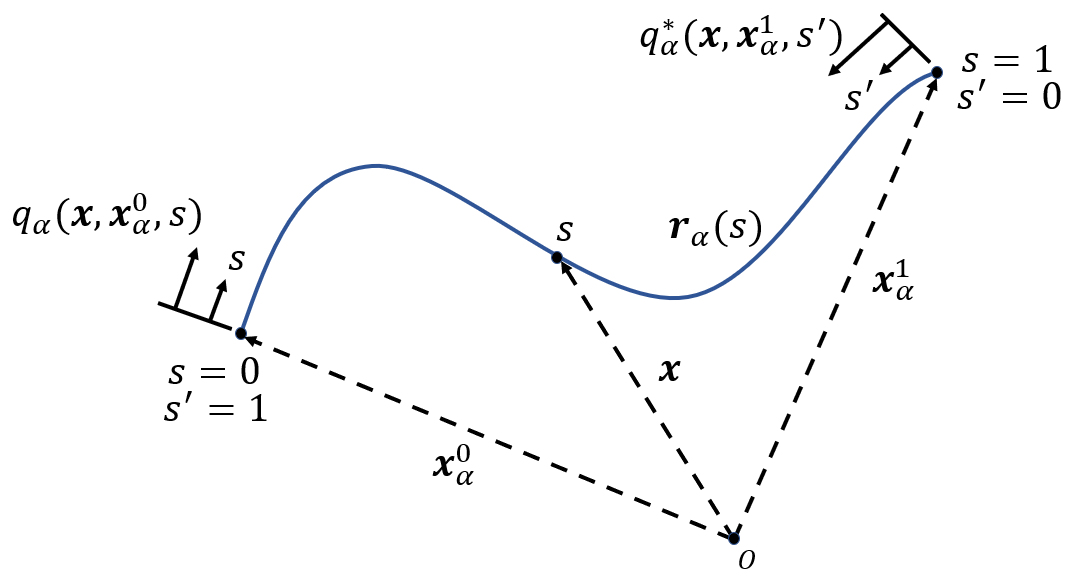}
	\caption{Single polymer chain fixed at both ends. }
	\label{fig:single_chain}
\end{figure}

We use the Continuous Gaussian Chain model for a single polymer chain \cite{fredrickson2006equilibrium}. 
In the undeformed state, the coarse-grained trajectory of the $\alpha$-th polymer chain is represented as a continuous 3-d space curve $\bfR_{\alpha}(s)$, where $s$ is the chain contour coordinate and varies along the chain contour, and is scaled such that $0 \leq s \leq 1$. 
The position vectors of the beginning and end points of the chain in the undeformed state are $\bfX_{\alpha}^0$ and $\bfX_{\alpha}^1$.

The chain is deformed under the deformation gradient $\bfF$. 
In the deformed state, $\bfr_{\alpha}(s)$ is a 3-d curve that represents the coarse-grained trajectory of the $\alpha$-th chain, as shown in Figure \ref{fig:single_chain}.
The position vectors of the beginning and end of the chain in the deformed state are $\bfx_{\alpha}^0$ and $\bfx_{\alpha}^1$.

Following \cite{treloar1975physics}, we use that the chain end-to-end vector is mapped under the macroscale deformation $\bfF$:
\begin{equation}\label{eq:end_to_end_vector_deformation}
    \bfx_{\alpha}^1- \bfx_{\alpha}^0=  \bfF  \left(\bfX_{\alpha}^1- \bfX_{\alpha}^0\right)
\end{equation}
    We note that the affine deformation assumption depends strongly on the assumption that there are no entanglements \cite{davidson2013nonaffine,davidson2016nonaffine,rubinstein1997nonaffine,rubinstein2002elasticity}. 

\subsubsection{ Partition Function and Average Segment Density}\label{sec: Partition Function and Average Segment Density}

Consider the $\alpha$-th chain that consists of $N$ coarse-grained polymer segments each of length $a$, and under the influence of a field $w(\bfx)$ that will be used to account for the excluded volume interactions \cite{fredrickson2002field}.

From \cite{fredrickson2006equilibrium}, the partition function, $Q_{\alpha}[w; \bfF]$, and the average segment density, $ \langle \hat{\rho}_{\alpha}(\bfx; \bfF) \rangle$, are:
\begin{align}
    \label{eq:Q_alpha_final}
    Q_{\alpha}[w;\bfF]=\frac{1}{V} \int \dm \bfx \ q_{\alpha}(\bfx, \bfx_{\alpha}^0, s) \ q_{\alpha}^*(\bfx, \bfx_{\alpha}^1, 1-s),
    \\
    \label{eq:rho_alpha_final}
    \langle \hat{\rho}_{\alpha}(\bfx; \bfF) \rangle = \frac{1}{V Q_{\alpha}[w; \bfF]}  \int\limits_0^1  \dm s \ q_{\alpha}(\bfx, \bfx_{\alpha}^0, s) \ q_{\alpha}^*(\bfx, \bfx_{\alpha}^1, 1-s).
\end{align}
Here, $ q_{\alpha}(\bfx, \bfx_{\alpha}^0, s) $ and $q_{\alpha}^*(\bfx, \bfx_{\alpha}^1, 1-s)$ are the partial partition functions for the two chain fragments, one from $0$ to $s$ and the other from $1$ to $s$, respectively, as shown in Figure \ref{fig:single_chain}.

$q_{\alpha}(\bfx, \bfx_{\alpha}^0, s) $ is obtained by solving the following PDE with the initial condition:
\begin{equation}\label{eq:q_pde}
    \frac{\partial q_{\alpha}(\bfx, \bfx_{\alpha}^0, s)}{\partial s}= \frac{a^2 N}{6} \nabla^2 q_{\alpha}(\bfx, \bfx_{\alpha}^0, s) - w(\bfx) q_{\alpha}(\bfx, \bfx_{\alpha}^0, s),
    \quad
    q_{\alpha}(\bfx, \bfx_{\alpha}^0, s)\Big|_{s=0}= (a N ^{1/2})^{3} \ \delta (\bfx- \bfx_{\alpha}^0).
\end{equation}

Similarly, $q_{\alpha}^*(\bfx, \bfx_{\alpha}^1, s')$ is obtained by solving the same PDE as in \eqref{eq:q_pde}, but with the initial condition corresponding to keeping the other end fixed:
\begin{equation}\label{eq:q_star_pde}
    \frac{\partial q_{\alpha}^*(\bfx, \bfx_{\alpha}^1, s')}{\partial s'}= \frac{a^2 N}{6} \nabla^2 q_{\alpha}^*(\bfx, \bfx_{\alpha}^1, s') - w(\bfx) q_{\alpha}^*(\bfx, \bfx_{\alpha}^1, s'),
    \quad
    q_{\alpha}^{*}(\bfx, \bfx_{\alpha}^1, s')\Big|_{s'=0}= (a N ^{1/2})^{3} \ \delta (\bfx- \bfx_{\alpha}^1).
\end{equation}

The initial conditions above correspond to the physical constraint that the beginning and end points of the $\alpha$-th chain are fixed at the given spatial positions $\bfx_{\alpha}^0$ and $\bfx_{\alpha}^1$, respectively.  

\subsubsection{Reduction to Classical Rubber Elasticity}\label{sec: Verification with the Classical Rubber Elasticity}

In the absence of excluded volume interactions, obtained by setting $w(\bfx )\equiv 0$, we can find closed-form solutions for $q_{\alpha}$ and $q_{\alpha}^{*}$:
\begin{align}
    \label{eq:q_analytical_soln}
    & q_{\alpha}(\bfx, \bfx_{\alpha}^0, s) = \left(\frac{3}{2\pi s}\right)^{3/2}  \exp \left( -\frac{3 |\bfx- \bfx_{\alpha}^0|^2}{2a^2N
 s} \right),
    \\
    \label{eq:q*_analytical_soln}
    & q_{\alpha}^{*}(\bfx, \bfx_{\alpha}^1, 1-s) = \left(\frac{3}{2\pi (1-s)}\right)^{3/2}  \exp\left( -\frac{3 |\bfx- \bfx_{\alpha}^1|^2}{2a^2N(1-s)} \right).
\end{align}
The partition function $Q_{\alpha}[w; \bfF]\Big|_{w=0}$ in \eqref{eq:Q_alpha_final} evaluates to the classical Gaussian distribution in 3-d:
\begin{equation}\label{eq:Q_alpha_no_field}
   Q_{\alpha}[w; \bfF]\Big|_{w=0} \propto \left(\frac{3}{\sqrt{\pi}} \right)^3   \left(\frac{a^2N}{6} \right)^{3/2}  \exp \left( -\frac{3}{2a^2N} |\bfx_{\alpha}^1- \bfx_{\alpha}^0|^2 \right).
\end{equation}

Because the chains do not interact, the free energy of the $\alpha$-th polymer chain, $H_{\alpha}$, is obtained from $Q_{\alpha}$ using $H_{\alpha}=-k_B T \log Q_{\alpha}$ to be:
\begin{equation}\label{eq:fe_alpha_no_field_with_F}
    H_{\alpha}[w; \bfF]\Big|_{w=0}=  \frac{1}{2}   \left( \frac{3k_B T}{a^2N} \right)  \left| \bfF  \left(\bfX_{\alpha}^1 - \bfX_{\alpha}^0\right) \right|^2 - \left( \frac{3k_BT}{2} \right).
\end{equation}

To account for the fact that chains are randomly oriented, we next average $H_{\alpha}[w;\bfF]\Big|_{w=0}$ over all possible orientations of the chain end-to-end vector by integrating \eqref{eq:fe_alpha_no_field_with_F} over all orientations.
That is, keeping $\bfF$ fixed, we integrate $\left(\bfX_{\alpha}^1 - \bfX_{\alpha}^0\right)$ over the sphere of appropriate radius.
The resulting expression for the orientationally-averaged free energy, $H^{avg}_{\alpha}[w; \bfF]\Big|_{w=0}$, is:
\begin{equation}\label{eq:W_alpha_avg}
    H^{avg}_{\alpha}[w; \bfF]\Big|_{w=0}= \frac{k_B T}{2} \left( \trace  (\bfF^T \bfF) - 3 \right).
\end{equation}
This result recovers the classical rubber elasticity result \cite{treloar1975physics}, also known as the incompressible Neo-Hookean elastic strain energy.

\subsection{Deformation of the Polymer Network}\label{sec:deformation of polymer network}

The pairwise excluded volume interactions are introduced through the field $w(\bfx)$ following \cite{fredrickson2002field}.
We introduce $\Bar{u}(|\bfx- \bfx'|)$, which is the pairwise interaction potential of mean force for two segments located at spatial coordinates $\bfx$ and $\bfx'$.
The corresponding partition function for the polymer network in the deformed state, $Z(\bfF)$ in the field-theoretic setting is:
\begin{equation}\label{eq:Z_afo_F_gen_final}
     Z(\bfF) \propto \int \Dm \rho  \int \Dm w  \  \exp{ \left( - \frac{H[w, \rho; \bfF]}{k_B T} \right)  },
\end{equation} 
where $H[w,\rho; \bfF]$ is the effective Hamiltonian of the polymer network, and has the expression:
\begin{equation}\label{eq:H_afo_w_rho_F_gen_final}
    \frac{H[w,\rho; \bfF]}{k_B T}   
    = 
    -  \int \dm \bfx \ w(\bfx) \rho(\bfx) 
    + \ \frac{1}{2 k_B T} \int \dm \bfx \int \dm \bfx'\ \rho(\bfx) \ \Bar{u}(|\bfx -\bfx'|) \ \rho(\bfx')      
    -\log \left( Q_1[w;\bfF] \ldots  Q_n[w;\bfF] \right) .
\end{equation}
The auxiliary fields $w(\bfx)$ and $\rho(\bfx)$ are interpreted as the fluctuating chemical potential field generated internally because of the inter-segment interactions and the fluctuating density of the polymer network, respectively \cite{fredrickson2002field}.
$Q_{\alpha}[w; \bfF]$ is the partition function for the $\alpha$-th chain in the polymer network under the influence of $w(\bfx)$, and is calculated using \eqref{eq:Q_alpha_final}.

The first term in \eqref{eq:H_afo_w_rho_F_gen_final} is the energy of interaction between the density and the chemical potential. The second term is the inter-segment interaction energy.
The third term is the entropic contribution due to chain stretching. 
The total Helmholtz free energy of polymer network in the deformed state, $H(\bfF)$, is evaluated from the partition function $Z(\bfF)$ in \eqref{eq:Z_afo_F_gen_final} using:
\begin{equation}\label{eq:H_afo_F}
    H(\bfF)= -k_B T \log Z(\bfF).
\end{equation}

In the deformed state, the average segment density of the polymer network, $\langle \hat{\rho}(\bfx; \bfF) \rangle$, is obtained as:
\begin{equation}\label{eq:avg_rho_hat_gen_final}
  \begin{split}
    \langle \hat{\rho}(\bfx; \bfF) \rangle \propto & \frac{1}{Z(\bfF)}  \int \Dm \rho  \int \Dm w  \
          \Bigg[ \exp \left( \int d\bfx \ w(\bfx) \rho(\bfx) -  \frac{1}{2 k_B T} \int \dm \bfx \int \dm \bfx' \rho(\bfx)   \Bar{u}(|\bfx -\bfx'|)  \rho(\bfx')   \right)  \\   
           & \left( \sum_{i=1}^n \left( \left(  \int\limits_0^1 \dm s \ q_i(\bfx, \bfx_i^{0}, s) \ q_i^*(\bfx, \bfx_i^{1}, 1-s)  \right)  \  \prod\limits_{\substack{k=1 \\ k\neq i}}^n Q_k[w;\bfF] \right) \right) \Bigg].
  \end{split}
\end{equation} 

\subsection{Strain Energy Density of the Polymer Network}\label{sec:strain energy density}

To obtain the continuum elastic response using nonlinear elasticity, we introduce the elastic energy density (per undeformed unit volume) $W(\bfF)$ and treat the RVE as a continuum material point.
This allows us to connect the total free energy $H(\bfF)$ evaluated for the RVE to $W(\bfF)$ at the corresponding spatial location:
\begin{equation}\label{eq:W_afo_H}
    W(\bfF)= \frac{H(\bfF)}{V} ,
\end{equation}
where $V$ is the volume of the RVE in the undeformed state.

We can then find the Piola-Kirchhoff stress tensor $\bfP$ and the fourth-order elasticity tensor $\mathbb{C}=[C_{ijkl}]$ using:
\begin{align}
    \label{eq:stress_def}
    & \bfP= \frac{\partial W}{\partial \bfF},
    \\
    \label{eq:elastic_moduli_def}
    & C_{ijkl}= \frac{\partial^2 W}{\partial F_{ij} \partial F_{kl}} \Bigg|_{\bfF = \bfI},
\end{align}
where $\bfI$ is the second-order identity tensor.

Applying \eqref{eq:stress_def}, and substituting from \eqref{eq:Z_afo_F_gen_final}, \eqref{eq:H_afo_w_rho_F_gen_final}, \eqref{eq:H_afo_F}, \eqref{eq:W_afo_H}, we have:
\begin{equation}\label{eq:piola_kirchhoff_stress_appendix}
  \begin{split}
\bfP = & \frac{\partial W}{\partial \bfF} 
     =  - \frac{k_B T}{V} \frac{1}{Z(\bfF)} \frac{\partial Z(\bfF)}{\partial \bfF} 
      = - \frac{k_B T}{V} \frac{1}{Z(\bfF)} \int \Dm \rho \int \Dm w \exp{ \left( - \frac{H[w, \rho; \bfF]}{k_B T} \right)  } \frac{\partial H[w,\rho; \bfF]}{\partial \bfF} \left( -\frac{1}{k_B T} \right) \\
      =  &- \frac{k_B T}{V} \frac{1}{Z(\bfF)} \int \Dm \rho \int \Dm w \exp{ \left( - \frac{H[w, \rho; \bfF]}{k_B T} \right)  } \sum\limits_{\alpha=1}^{n} \frac{\partial}{\partial \bfF}\log Q_{\alpha}[w;\bfF]. \\
  \end{split}
\end{equation}

Then, defining the stress operator for the $\alpha$-th chain, $\hat{\bfP}_{\alpha}$, as:
\begin{equation}\label{eq:stress_operator_singe_chain}
    \hat{\bfP}_{\alpha}:= -\frac{\partial \ }{\partial \bfF} \log Q_{\alpha}[w;\bfF],
\end{equation}
we can write $\bfP$ as the statistical average \cite[Section 4.1.3]{fredrickson2006equilibrium} of $\hat{\bfP}_{\alpha}$:
\begin{equation}\label{eq:stress_afo_stress_operator}
    \bfP 
    =
    \frac{k_B T}{V} \left( \frac{1}{Z(\bfF)} \int \Dm \rho \int \Dm w \exp{ \left( - \frac{H[w, \rho; \bfF]}{k_B T} \right)  } \sum\limits_{\alpha=1}^{n} \hat{\bfP}_{\alpha} \right)
    =
    \frac{k_B T}{V} \sum\limits_{\alpha=1}^{n} \langle \hat{\bfP}_{\alpha} \rangle.
\end{equation}

\subsection{Mean-field Assumption}\label{sec: mean-field Assumption}

The functional integration over the fields $w$ and $\rho$ in \eqref{eq:Z_afo_F_gen_final} and \eqref{eq:avg_rho_hat_gen_final} makes it expensive to evaluate $H(\bfF)$ and $\langle \hat{\rho}(\bfx; \bfF) \rangle$.
Therefore, it is common to use a mean-field assumption to simplify the functional integration in the expression for $Z(\bfF)$ in \eqref{eq:Z_afo_F_gen_final} \cite{fredrickson2006equilibrium}.
This assumption implies that functional integration over the fields $w$ and $\rho$ is dominated by the mean-fields $\Bar{w}$ and $\Bar{\rho}$, respectively. 
The mean-fields $\Bar{w}$ and $\Bar{\rho}$ are obtained by requiring the effective Hamiltonian $H[w, \rho; \bfF]$ in \eqref{eq:H_afo_w_rho_F_gen_final} to be  stationary with respect to variations in $w(\bfx)$ and $\rho(\bfx)$. This gives the self-consistent mean-field conditions:
\begin{align}
    \label{eq:w_mf_condition}
    & \frac{\delta H[w, \rho; \bfF]}{\delta w}\Big|_{w=\bar{w}}=0,
    \\
    \label{eq:rho_mf_condition}
    & \frac{\delta H[w, \rho; \bfF]}{\delta \rho}\Big|_{\rho=\bar{\rho}}=0.
\end{align}
Using the mean-field assumption, $Z(\bfF)$ in \eqref{eq:Z_afo_F_gen_final} simplifies to:
\begin{equation}\label{eq:Z_n_chains_mf_1}
      Z(\bfF)\ \approx  \exp \left( -\frac{H[\Bar{w}, \Bar{\rho}; \bfF]}{k_B T} \right),
\end{equation}     
where, $H[\Bar{w}, \Bar{\rho}; \bfF]$ is the effective Hamiltonian in \eqref{eq:H_afo_w_rho_F_gen_final} evaluated using the mean-fields $\bar{w}$ and $\bar{\rho}$. 
Using \eqref{eq:H_afo_F} and \eqref{eq:Z_n_chains_mf_1}, the total free energy of polymer network, $H(\bfF)$ under the mean-field assumption is:
\begin{equation}\label{eq:H_afo_F_final}
    \frac{H(\bfF)}{k_B T}  
    =   
    -\int \dm \bfx \ \Bar{w}(\bfx)  \Bar{\rho}(\bfx) \ 
    + \ \frac{1}{2 k_B T} \int \dm \bfx \int \dm \bfx' \ \Bar{\rho}(\bfx) \ \Bar{u}( | \bfx- \bfx'| ) \ \Bar{\rho}(\bfx') 
    - \log \left(  Q_{1}[\Bar{w}; \bfF] \cdots Q_{n}[\Bar{w}; \bfF]  \right) .
\end{equation}
Further, the average segment density, $\langle{\Hat{\rho}}(\bfx; \bfF)\rangle$, in \eqref{eq:avg_rho_hat_gen_final} simplifies to:
\begin{equation}\label{eq:avg_rho_hat_final_mf}
    \langle\hat{\rho}(\bfx; \bfF)\rangle \approx \sum\limits_{\alpha=1}^{n} \langle\hat{\rho}_{\alpha}(\bfx; \bfF)\rangle \Big|_{w=\Bar{w}} ,
\end{equation} 
where $\langle\Hat{\rho}_{\alpha}(\bfx; \bfF)\rangle$ is the average segment density of the $\alpha$-th chain in the polymer network, obtained using \eqref{eq:rho_alpha_final}.

\subsection{Excluded Volume Interaction}\label{sec: Inter-Segment Interaction: Excluded Volume Approach}

Polymer segments in the network are considered to interact with each other according to a pairwise interaction potential of mean force $\bar{u}$ whose physical origin is due to excluded volume effects.
We account for the excluded volume effects by modeling a pairwise inter-segment interaction using a simple Dirac delta potential of mean force \cite{zimm1953excluded,doi1986theory}:
\begin{equation}\label{eq:dirac_delta_potential_excluded_volume} 
   \Bar{u}(|\bfx- \bfx'|)=  k_B T \ u_0 \ \delta(|\bfx - \bfx'|).
\end{equation} 
where $u_0$ is the excluded volume parameter.
This form of inter-segment interaction potential assumes the presence of a solvent in the polymer network system with low density \cite{deam1976theory, doi1986theory}. The solvent mediates the interactions among polymer segments.
For $u_0>0$, implying repulsion between the segments, the excluded volume potential $\bar{u}$ in \eqref{eq:dirac_delta_potential_excluded_volume} is positive-definite and has an inverse; following \cite{fredrickson2006equilibrium}, this simplifies the field theory equations in \eqref{eq:Z_afo_F_gen_final} to:
\begin{equation}\label{eq:Z_simplified}
    Z(\bfF) \propto \int \Dm w \exp \left(-\frac{H[w; \bfF]}{k_B T} \right),
\end{equation}
where $H[w;\bfF]$ is the effective Hamiltonian of polymer network in the simplified field theory:
\begin{equation}\label{eq:H_simplified}
    \frac{H[w;\bfF]}{k_B T}= \frac{1}{2u_0}  \int \dm \bfx \ (w(\bfx))^2  - \log (Q_1[w; \bfF]  \cdots  Q_n[w; \bfF] ).
\end{equation}
Equations \eqref{eq:Z_simplified} and \eqref{eq:H_simplified} present the simplified field theory for the deformation of polymer network that is used in this work. 
The partition function in \eqref{eq:Z_simplified} is evaluated using the mean-field assumption as:
\begin{equation}\label{eq:Z_simplified_mf}
    Z(\bfF) \approx  \exp \left( -\frac{H[\Bar{w}; \bfF]}{k_B T} \right),
\end{equation}
where $H[\bar{w};\bfF]$ is the effective Hamiltonian in \eqref{eq:H_simplified} evaluated using the mean-field $\Bar{w}$. 
The mean-field $\Bar{w}$ is obtained by solving the stationarity condition for the effective Hamiltonian $H[w;\bfF]$:
\begin{equation}\label{eq:H_simplified_mf_condition}
    \frac{\delta H[w; \bfF]}{\delta w}\Big|_{w=\Bar{w}} =0.
\end{equation}

For the assumed form of the excluded volume interaction potential as in \eqref{eq:dirac_delta_potential_excluded_volume}, there is alternatively an expression for the average segment density \cite{fredrickson2006equilibrium}:
\begin{equation}\label{eq:rho_alternate}
   \langle \hat{\rho}(\bfx; \bfF) \rangle = \frac{1}{u_0} \langle w(\bfx) \rangle = \frac{\bar{w}(\bfx)}{u_0}
\end{equation}
where $\langle w(\bfx) \rangle$ is the statistical average of the fluctuating field $w(\bfx)$, and we use that under the mean-field assumption $\langle w(\bfx) \rangle= \Bar{w}(\bfx)$.
Finally, using \eqref{eq:Z_simplified_mf} and \eqref{eq:H_afo_F}, we obtain the total free energy of polymer network, $H(\bfF)$ for the simplified field theory as:
\begin{equation}
    \frac{H(\bfF)}{k_B T}= \frac{1}{2u_0} \int \dm \bfx (\Bar{w}(\bfx))^2 - \log (Q_1[\Bar{w}; \bfF]  \cdots  Q_n[\Bar{w}; \bfF] ) ,
\end{equation}
where $\Bar{w}(\bfx)$ is obtained by self-consistently solving \eqref{eq:rho_alternate} and the mean-field condition in \eqref{eq:H_simplified_mf_condition}. 

\subsection{Representative Volume Element Averaging: 8-Chain Model}\label{sec: Chain Averaging: $8$ Chain Model}

A typical polymer network consists of a large number of cross-linked polymer chains (Fig. \ref{fig:8-chain}) with random orientations at each continuum point, and is very challenging to directly solve.
To simplify this problem, we adopt the 8-chain model for the RVE \cite{arruda1993three}.
The 3-d RVE is assumed to be a cube in the undeformed configuration, with 8 chains running between the center and each of the corners.
The cube is assumed to be oriented along the principal directions of the macroscale right stretch tensor $\bfU$, where $\bfU$ is the tensor square root of the deformation $\bfF$ or alternatively is the positive-definite part of the right polar decomposition of $\bfF$.

We assume that each chain begins ($s=0$) at the center of the cube which is also taken to be the origin, and the chains terminate ($s=1$) at the corners.
Denoting the terminating point of the chains in the undeformed and deformed state, respectively, by $\bfX_1^1, \ldots \bfX_8^1$ and $\bfx_1^1, \ldots \bfx_8^1$, the relation between the end-to-end vectors in the undeformed and deformed configurations from \eqref{eq:end_to_end_vector_deformation} is:
\begin{equation}
   \bfx_{\alpha}^1= \bfF \ \bfX_{\alpha}^1, \qquad \alpha=1, \ldots 8 .
\end{equation}
For a given value of $\bfF$, the right stretch tensor $\bfU$ is used to orient the cube, and the equation above provides the initial conditions for the partial partition functions $q$ and $q^*$ of each chain in \eqref{eq:q_pde} and \eqref{eq:q_star_pde}.

\section{Numerical Method}\label{sec:numerical method}

Since the model with excluded volume interactions is not amenable to simple closed-form solutions, we turn to numerical solutions.
The goal is to evaluate the total free energy $H(\bfF)$ and average segment density $\langle \Hat{\rho}(\bfx; \bfF)  \rangle$.
Our numerical method has the following steps:
\begin{enumerate}
    \item We first generate an initial field $w(\bfx)=w^0(\bfx)$. The initial guess can be based on heuristics when possible.

    \item We next solve for the average segment density $\langle \hat{\rho}(\bfx; \bfF) \rangle$ -- using $q$ and $q^*$ obtained by solving \eqref{eq:q_pde} and \eqref{eq:q_star_pde} with the given $w(\bfx)$ -- and the total free energy $H(\bfF)$ using \eqref{eq:H_simplified_mf_condition}.
    
    \item The field $w$ is updated using \eqref{eq:rho_alternate} as $w(\bfx)= u_0 \langle \Hat{\rho}(\bfx; \bfF) \rangle $.

    \item In turn, we update $\langle \hat{\rho}(\bfx; \bfF) \rangle$ and $H(\bfF)$ as above.

\end{enumerate}
This iteration continues until we reach convergence, which we define as a relative change in the total free energy of less than $0.1 \%$.

\begin{figure*}[ht!]
	\subfloat[]{\includegraphics[width=0.37\textwidth]{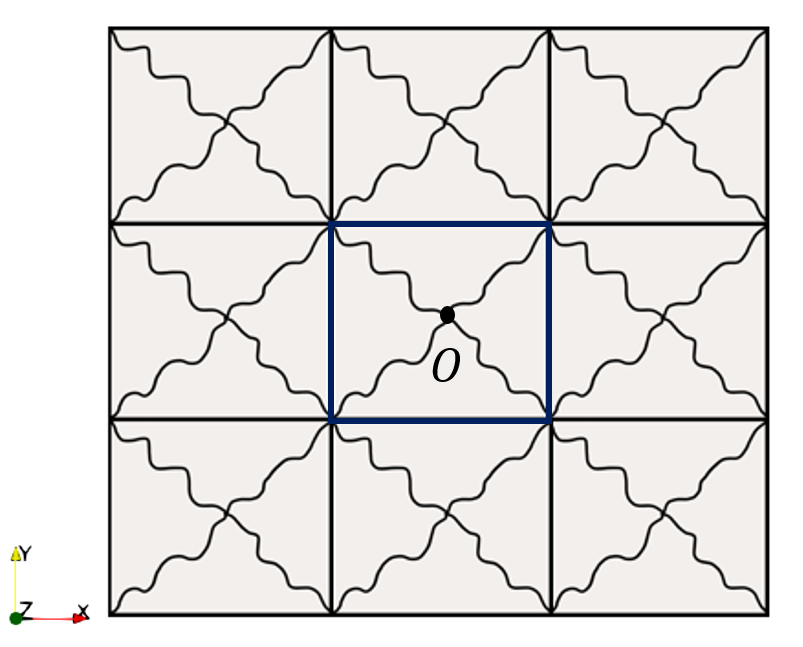}}
	\hfill
	\subfloat[]{\includegraphics[width=0.57\textwidth]{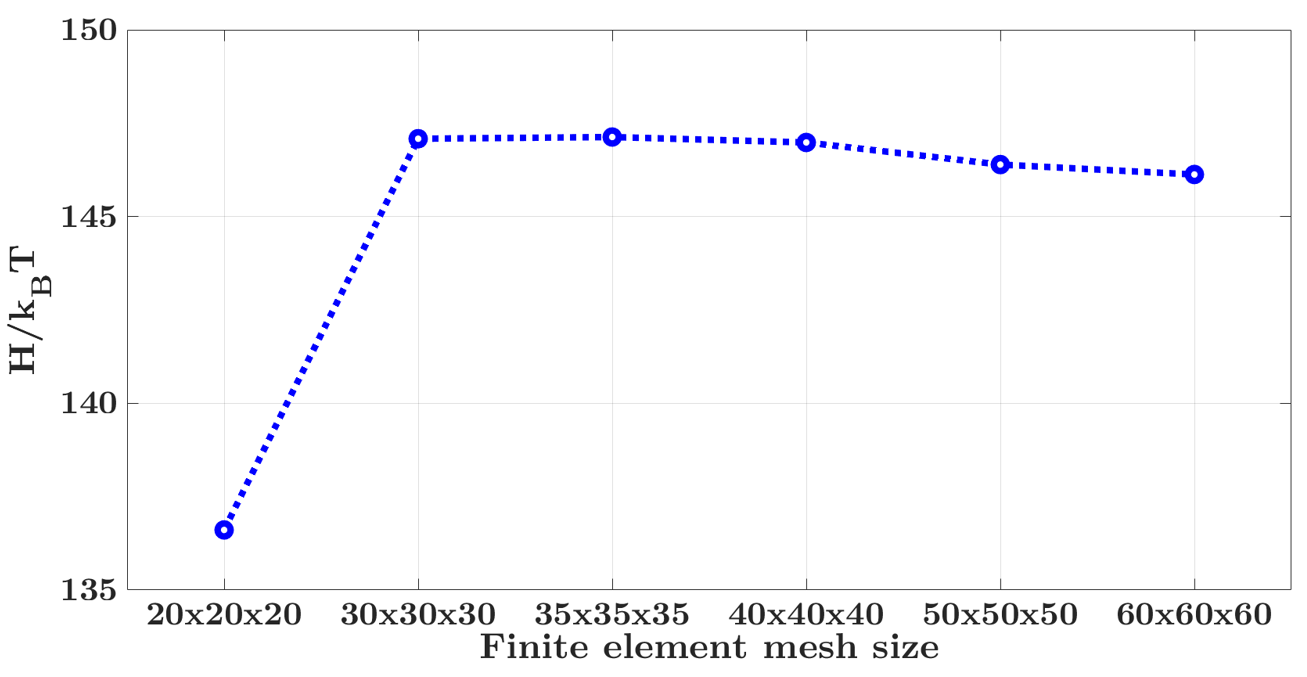}}
	\caption{
	    (a) Front view schematic of a bigger physical domain in 3-d that consists of 27 RVEs, each with $8$ chains. The central $8$ chain RVE (highlighted) is used for the free energy calculation. 
	    (b) The convergence of the energy as the mesh is refined, shown by plotting the free energy of a single chain without external field as a function of mesh size. The converged mesh size of $35^3$ elements is used for the numerical computations.
	Note that the saddle point nature of the problem can lead to non-monotonic convergence.
    }
    \label{fig:numerics}
\end{figure*}

For the solution of $q$ and $q^*$ in \eqref{eq:q_pde} and \eqref{eq:q_star_pde}, we use the finite element method (FEM) in the open-source FEniCS framework \cite{langtangen2017solving}. 
The spatial domain is discretized using first-order Lagrange family finite elements.
The integration along the chain with respect to $s$ in \eqref{eq:q_pde} and in \eqref{eq:q_star_pde} is performed using the implicit Crank-Nicholson finite difference method with $100$ steps for $s \in (0,1)$.
We test convergence of the FEM discretization as in Figure \ref{fig:numerics}(b).

While the RVE averaging nominally requires only 8 chains, these chains interact not only with each other, but also with other chains that are not contained in the RVE. 
To account for this, we use periodic images of the RVE; we find that one image on each face of the cubic RVE is sufficient, giving us 27 cubes over which we must perform various integrations; Figure \ref{fig:numerics}(a) shows a schematic projection of this in 2-d, where the highlighted central RVE is used for the energy calculations.
When the deformation is applied, the image RVEs are deformed following the central RVE.
Similarly, $w(\bfx)$ is defined over the larger cluster of RVEs for performing, but need only be solved on the central RVE using periodicity.

\section{Elastic Response with Excluded Volume Interactions}\label{sec: results and discussion}

In the calculations reported here, we use the following model parameters: total chain contour length $L=0.12\unit{\micro\meter}$, number of polymer segments in single chain $N=100$, excluded volume parameter $u_0=0.005
 \ v_{seg}$ where $v_{seg}=a^3$ is the volume of an individual monomer segment, and temperature $T=303\unit{\kelvin}$. 
    This choice of $u_0$ gives an excluded cube with side $0.17 a$, and corresponds to the Flory-Huggins interaction parameter $\chi=0.4975$, using $u_0= (1-2 \chi) v_{seg}$.
    This characterizes a good solvent that is very close to the $\Theta$-point \cite{de1969some,fredrickson2006equilibrium,milner2009chi,nistane2022estimation}.

For the numerical calculations, larger values of $u_0$ in 3-d require an excessively fine mesh to converge; however, the calculations described below qualitatively agree with 2-d calculations -- where much larger values of $u_0$ can be used -- and with a few representative 3-d calculations that were conducted with a larger value of $u_0$ and a very fine mesh.

\subsection{Identifying the Stress-Free Free Energy Minimizing State}

We first note that without excluded volume effects, all models based on the Gaussian chain approximation would predict that the polymer network shrinks to a point ($\bfF = {\bf 0}$) because that would maximize the configurational entropy of the chain.
Therefore, as we add excluded volume effects, we find that the equilibrium -- stress-free or minimum energy -- volume of the polymer network RVE increases.
    Specifically, the equilibrium state is achieved as a balance between the competing effects of entropic shrinkage and excluded volume repulsion between monomers; \cite{wittmer2007intramolecular, lang2015conformations} examine related issues in greater depth.

\begin{figure}[ht!]
	\centering
    \includegraphics[width=0.95\textwidth]{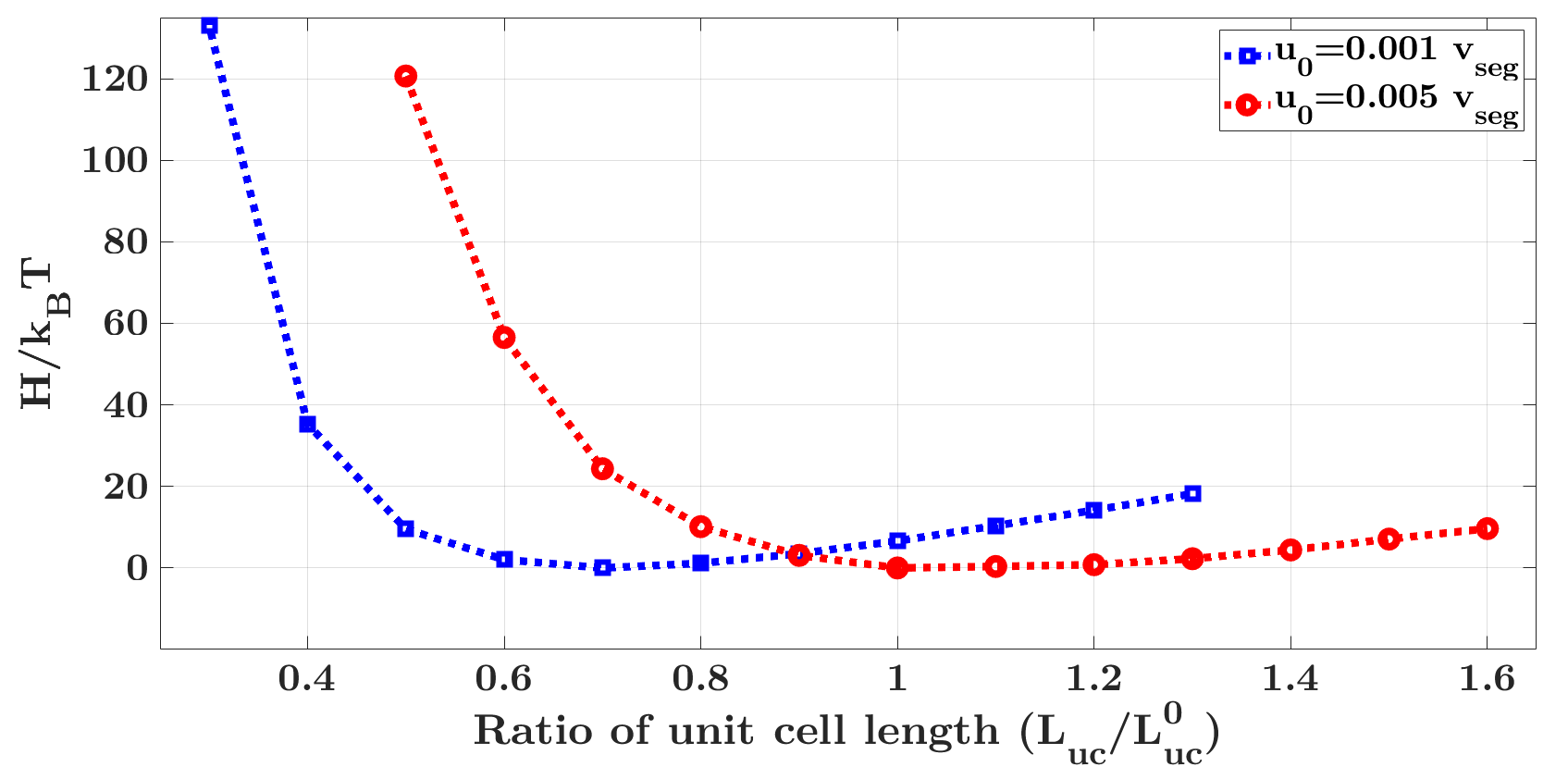}
	\caption[Free energy vs ratio of unit cell length ]{The total free energy as a function of the relative stretch for $u_0=0.001 \ v_{seg}$ and $u_0=0.005 \ v_{seg}$.}
    \label{fig:H_vs_unit_cell_length}
\end{figure}

We denote the initial side of the cubic RVE by $L_{uc}^0:= 2aN^{1/2}/\sqrt{3}$, where $aN^{1/2}$ is the RMS average diameter of an unconstrained chain, and the pre-factor accounts for the geometry of the chain aligned along the diagonal of the RVE.
However, we emphasize that this is \textit{not} the free energy minimizing state, i.e., it is not stress-free, in the presence of excluded volume interactions.
To this initial state, we apply a deformation of the form $\bfF = \lambda \bfI$, and compute the free energy for various values of $\lambda$.
Figure \ref{fig:H_vs_unit_cell_length} shows the total free energy $H$ as a function of $\lambda = L_{uc}/L_{uc}^0$ for $u_0=0.001 \ v_{seg}$ and $u_0=0.005 \ v_{seg}$.
We find that the stress-free stretches, i.e. the free-energy minimizing stretches, for $u_0=0.001 \ v_{seg}$ and $u_0=0.005 \ v_{seg}$ are respectively at $\lambda =0.7$ and $\lambda=1$ respectively. 
The large increase in the equilibrium volume of the polymer network system with an increase in the excluded volume parameter $u_0$ is consistent with the phenomena of equilibrium swelling for polymeric gels \cite{hong2008theory, kamata2015design, tanaka1979kinetics, jagur2010polymeric}.

In all of the subsequent calculations discussed below, we set the undeformed state to correspond to the free energy minimizing stress-free state.

\subsection{Volumetric and Shear Response, and Near-Incompressibility}

We use the strain energy density $W(\bfF)$ to obtain the mesoscale elastic response using nonlinear elasticity; specifically, \eqref{eq:stress_def} and \eqref{eq:elastic_moduli_def} are used to obtain the stress-stretch response and the elastic moduli respectively. 
We assume below that the network can be treated as approximately isotropic despite the 8-chain model.

To obtain the bulk and shear moduli, we impose deformations of the form:
\begin{equation}\label{eq:deformation_gradients}
    \bfF_v = 
        \begin{bmatrix}
        \lambda & 0 & 0 \\
        0 & \lambda & 0 \\
        0 & 0 & \lambda
        \end{bmatrix}
        \quad , \quad
    \bfF_s = 
        \begin{bmatrix}
        1 & \kappa_s & 0 \\
        0 & 1 & 0 \\
        0 & 0 & 1
        \end{bmatrix}.
\end{equation}

The bulk and shear moduli, $K$ and $G$, respectively can be computed using:
\begin{equation}\label{eq:elastic_moduli}
    K =\frac{1}{9}\frac{\partial^2W}{\partial \lambda^2}\Big|_{\lambda\approx 1} \quad , \quad
    G=\frac{\partial^2W}{\partial \kappa_s^2}\Big|_{\kappa_s \approx 0}. 
\end{equation}
We find $K=52.06\unit{\kilo\pascal}$ and $G=0.60 \unit{\kilo\pascal}$.
Using isotropic linearized elasticity, this gives the Poisson's ratio $\nu= \frac{3K-2G}{6K+2G} = 0.4943$ and the elastic modulus $E=\frac{9KG}{3K+G} =1.79 \unit{\kilo\pascal}$.
These elastic moduli are consistent with polymer network gels \cite{norman2021measuring, feig2018mechanically, ohm2021electrically, xu2023concurrent, coyle2018bio, li2014stiff, sun2022rheology, cai2011periodic, shan2015rigidity}.
We highlight that $K$ is 2 orders of magnitude larger than $G$, and $\nu$ is very close to the incompressible limit of $0.5$.

\begin{figure*}[ht!]
    \includegraphics[width=0.94\textwidth]{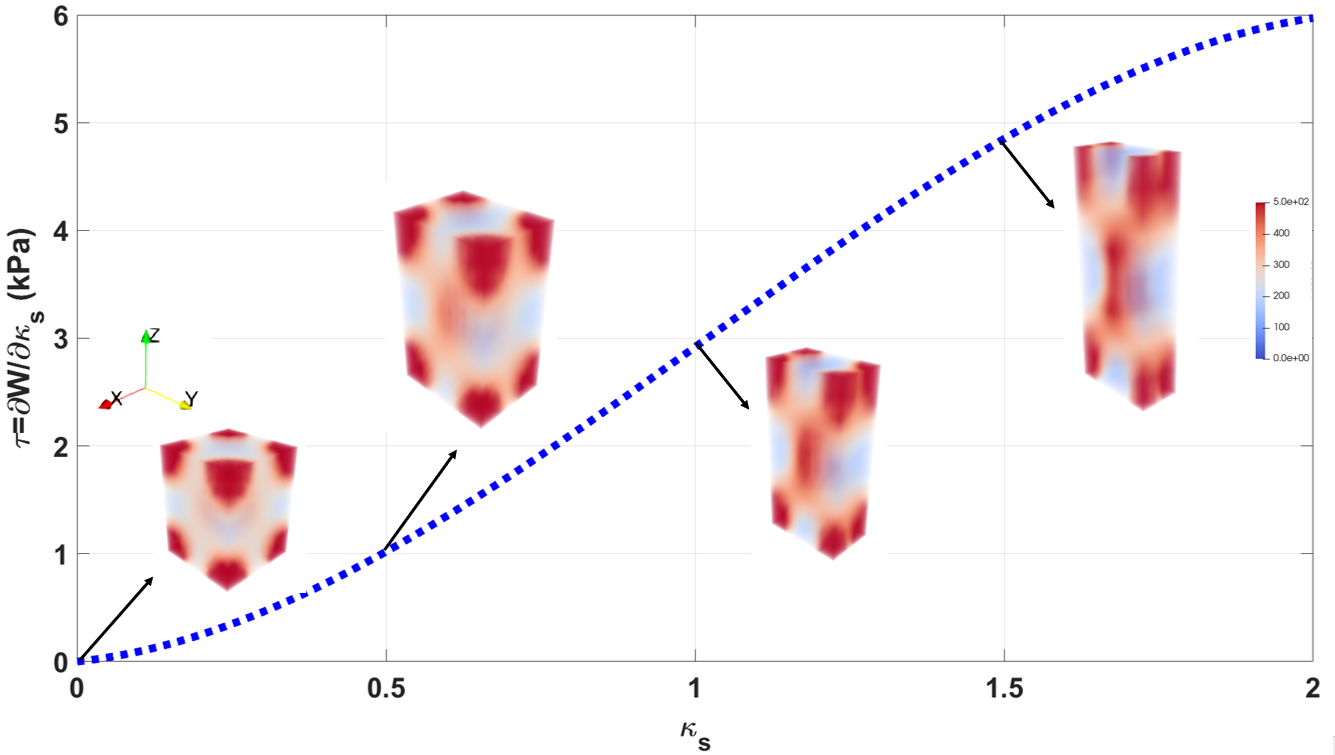}
    \caption{Shear stress $\tau$ vs. shear strain $\kappa_s$ in simple shear. The segment density for the RVE at various stretches are shown in the insets. We notice that the chains have higher concentrations at the ends when the deformation is small, but are more uniformly distributed as the deformation increases. Note that the RVE itself does not appear to be sheared because the chain-averaging approach aligns the averaging RVE along the principal directions that correspond to the maximum and minimum elongation directions (Fig. \ref{fig:8-chain}).}
    \label{fig:shear_stres_with_density}
\end{figure*}

We next examine the shear stress vs. shear strain curve.
In principle, the shear stress can be computed using $\tau= \frac{\partial W}{\partial \kappa_s}$.
To avoid a lot of noise from numerical differentiation, we fit $W$ by a polynomial and then differentiate the polynomial to obtain the curve shown in Figure \ref{fig:shear_stres_with_density}.

\subsection{Extensional Response: Emergent Strain-Softening and Strain-Stiffening}

\begin{figure*}[htb!]
    \includegraphics[width=0.94\textwidth]{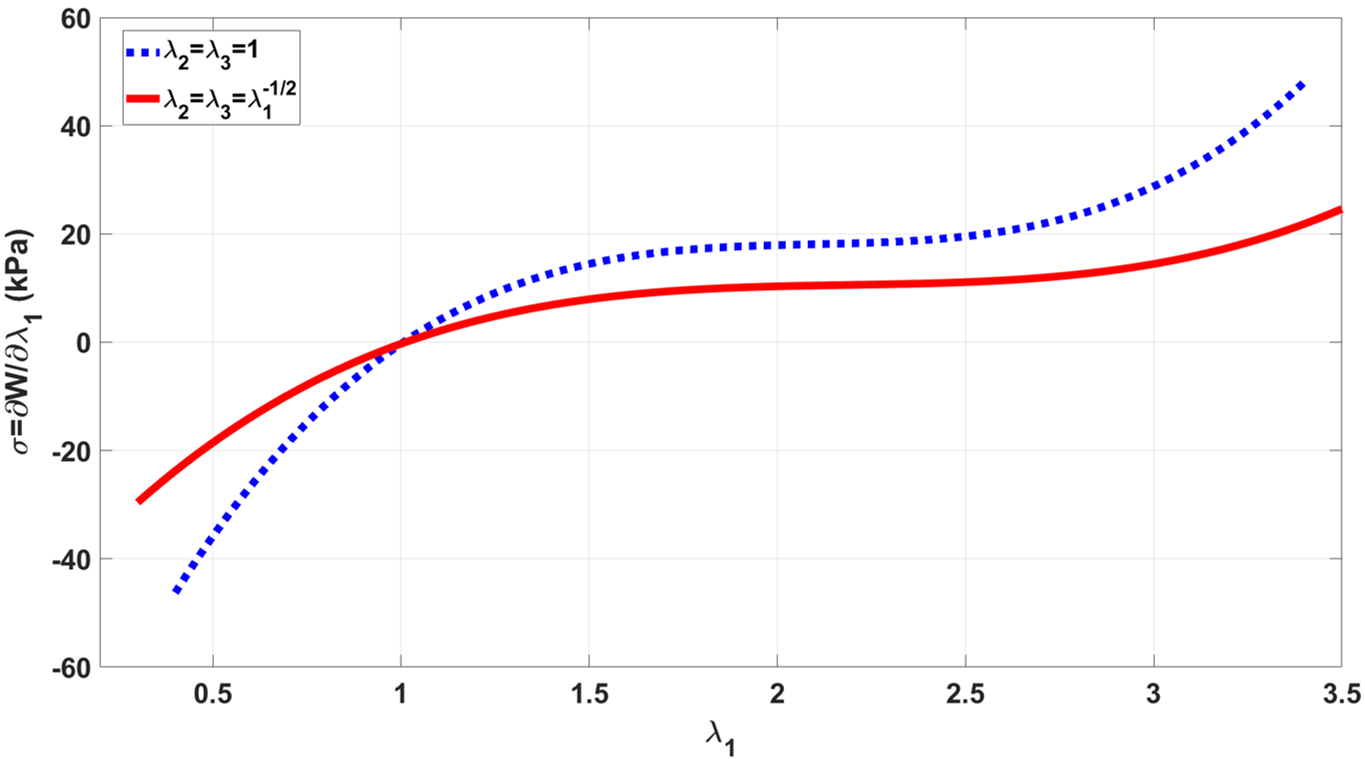}
    \caption{The stress-strain curves for the constrained and volume-preserving extensional loadings. We notice that the stresses and the tangent moduli are both significantly larger when the system is constrained to undergo deformations that do not preserve the volume.}
    \label{fig:extension-compare}
\end{figure*}

\begin{figure*}[htb!]
    \includegraphics[width=0.94\textwidth]{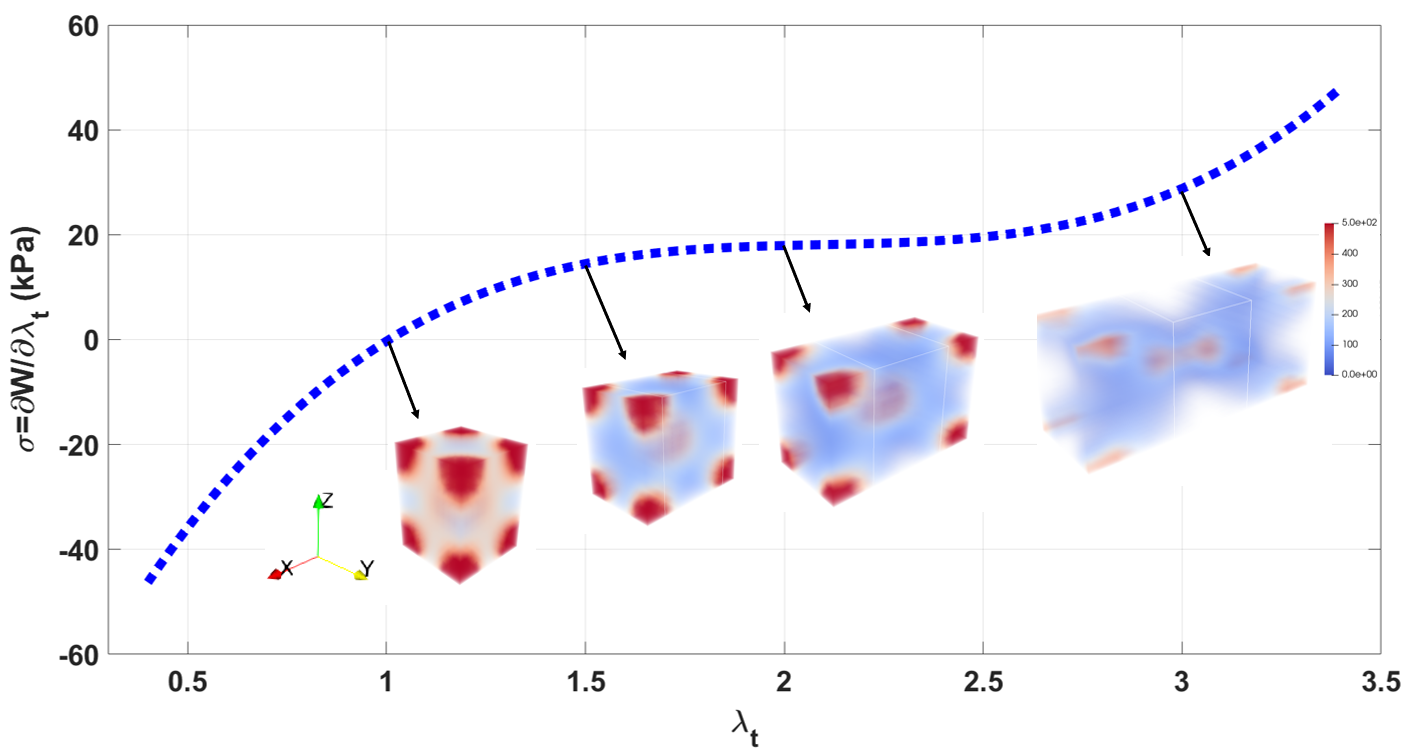}
    \caption{Extensional stress $\sigma$ vs. extensional stretch $\lambda_t$ for the constrained case from Figure \ref{fig:extension-compare}. The segment density for the RVE at various stretches are shown in the insets. As with shearing, we notice that the chains have higher concentrations at the ends when the deformation is small, but are more uniformly distributed as the deformation increases.}
    \label{fig:normal_stres_with_density}
\end{figure*}

We next examine extensional loading where the deformation has the form:
\begin{equation}
    \bfF_e = 
        \begin{bmatrix}
        \lambda_1 & 0 & 0 \\
        0 & \lambda_2 & 0 \\
        0 & 0 & \lambda_3
        \end{bmatrix}
\end{equation}
Here, we consider $\lambda_1$ as the extensional stretch of interest.
We make 2 different choices for the transverse stretches $\lambda_2$ and $\lambda_3$: the ``constrained case'' where they are constrained such that $\lambda_2=\lambda_3=1$; and the ``volume-preserving case'' where they are set to be volume-preserving such that $\lambda_2 = \lambda_3 = \lambda_1^{-\half}$.
Note that the second case is approximately equivalent to having no transverse stress.
To obtain the extensional stress, we use $\sigma= \frac{\partial W}{\partial \lambda_1}$.
Figure \ref{fig:extension-compare} compares the stress-strain response of these cases.
We notice that the constrained case has significantly higher stresses and tangent moduli.
Figure \ref{fig:normal_stres_with_density} shows the evolution of the chain density with stretch for the constrained case.

\begin{figure*}[ht!]
    \subfloat[Free energy density $W$ vs stretch $\lambda_1$, with  $\lambda_2=\lambda_3=1$]{\includegraphics[width=0.48\textwidth]{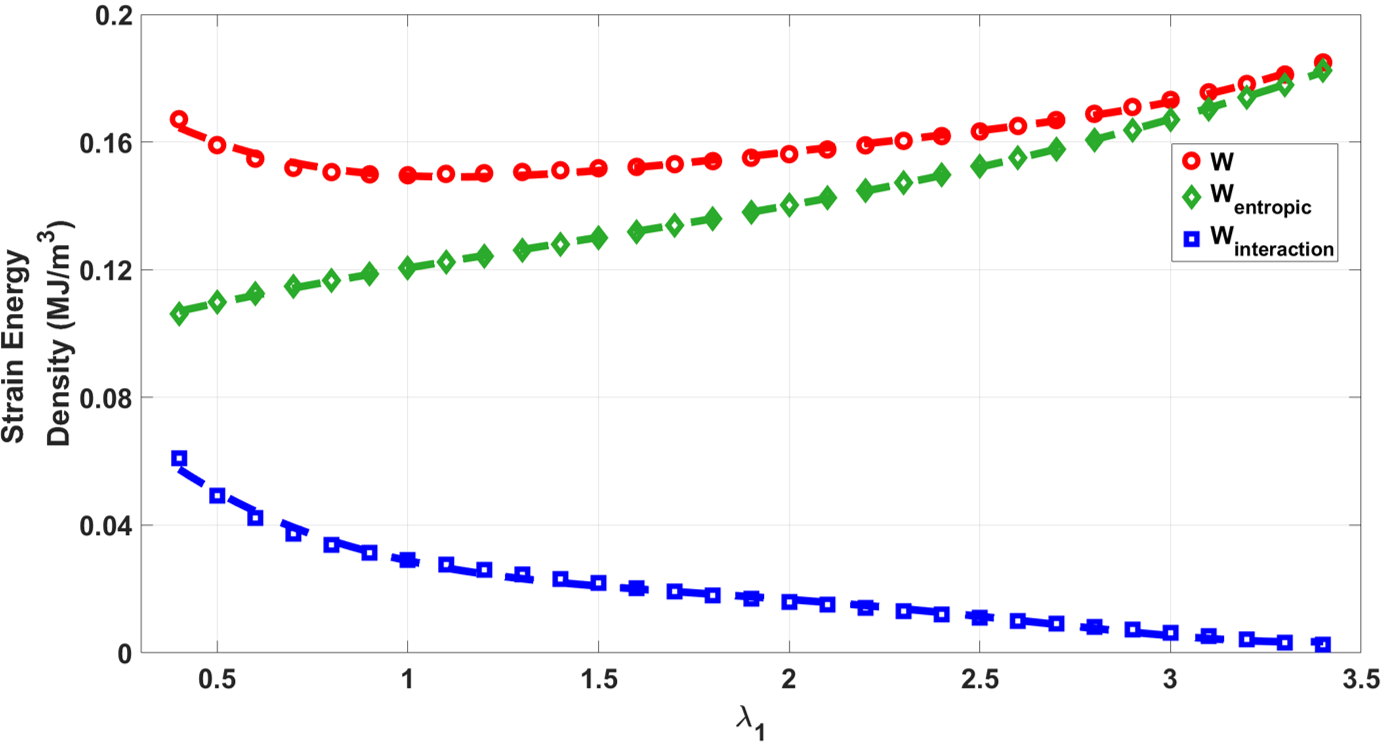}}
    \hfill
    \subfloat[Free energy density $W$ vs stretch $\lambda_1$ with $\lambda_2=\lambda_3=\lambda_1^{-1/2}$]{\includegraphics[width=0.48\textwidth]{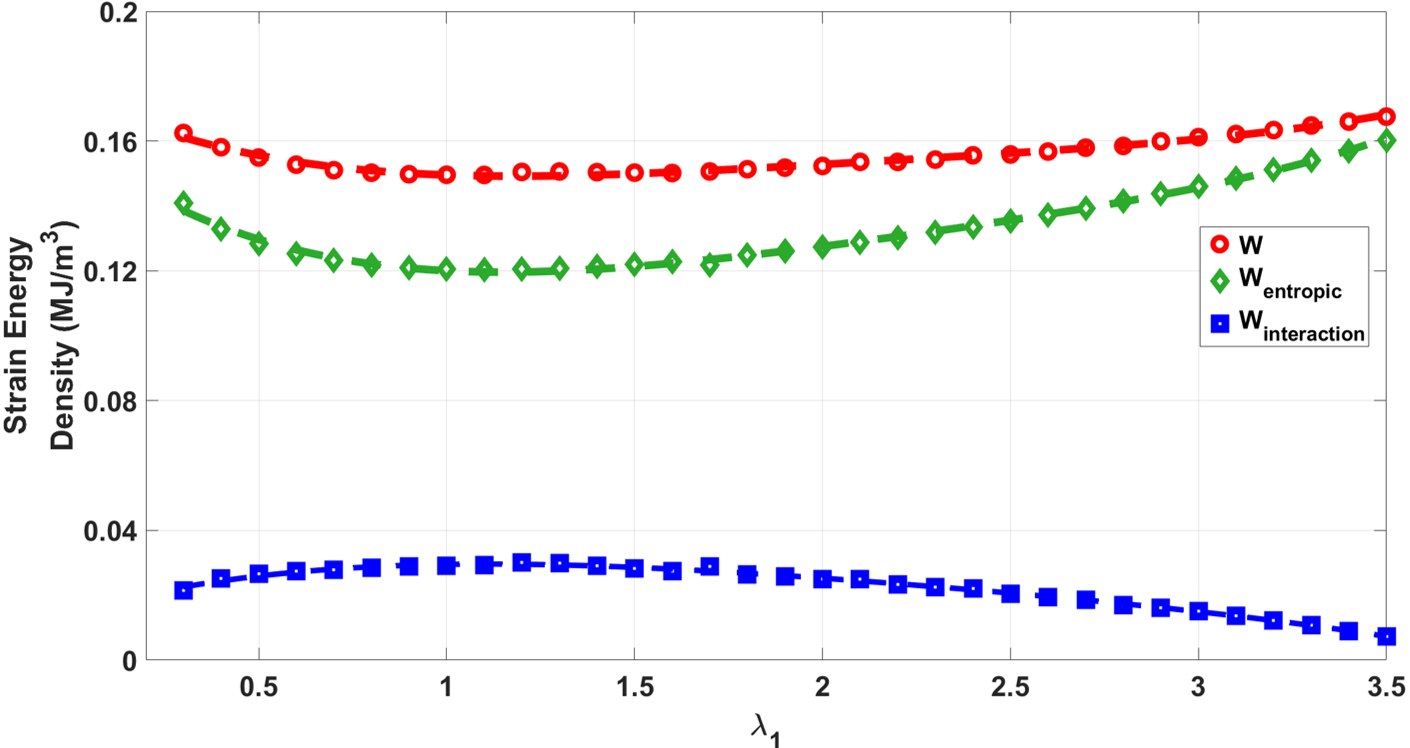}}
    \caption{The free energy $W$ is decomposed into entropic $W_{entropic}$ and excluded volume interaction $W_{interaction}$ contributions for the constrained and volume-preserving cases. The symbols show the simulation results, and the lines show best fits that are differentiated to obtain stress-strain curves.}
    \label{fig:entropy-interaction-decomposition}
\end{figure*}

From Figures \ref{fig:extension-compare} and \ref{fig:normal_stres_with_density}, we notice that both cases show a pronounced strain-softening and strain-stiffening behavior that is characteristic of many real polymer networks such as polymeric gels.
However, Gaussian chains do not show such behavior, and it is typical to use chains with limiting extensibility -- such as the inverse Langevin approximation -- to model this behavior.
Here, we find that it is a consequence of the competition between the excluded volume parameter and the entropy.
Figure \ref{fig:entropy-interaction-decomposition} shows the decomposition of the free energy $W$ into entropic $W_{entropic}$ and excluded volume interaction $W_{interaction}$ contributions.
We observe that the excluded volume contribution is less than the entropic contribution in both cases.
Further, we notice that $W_{entropic}$ monotonically increases with $\lambda_1 > 1$, and is consistent with the stretching of the Gaussian polymer chains; however, $W_{interaction}$ monotonically decreases with $\lambda_1 > 1$, and is consistent with the chains being more oriented, and hence having fewer excluded volume interactions.
We notice that in the approximate
 range $1.5<\lambda_1<2.5$ where we see strain softening, the decrease in the excluded volume interaction is faster than the rise in the entropic contribution, causing softening.
For $\lambda_1>3$, we have the opposite trend in that the entropy increases faster than the decrease in the excluded volume interaction, causing stiffening.
In summary, strain softening occurs because of the initial decrease in excluded volume interactions, and subsequent strain stiffening occurs because of the later increase in entropic effects.

\subsection{Effect of Chain Length} 

\begin{figure*}[htb!]
    \includegraphics[width=0.94\textwidth]{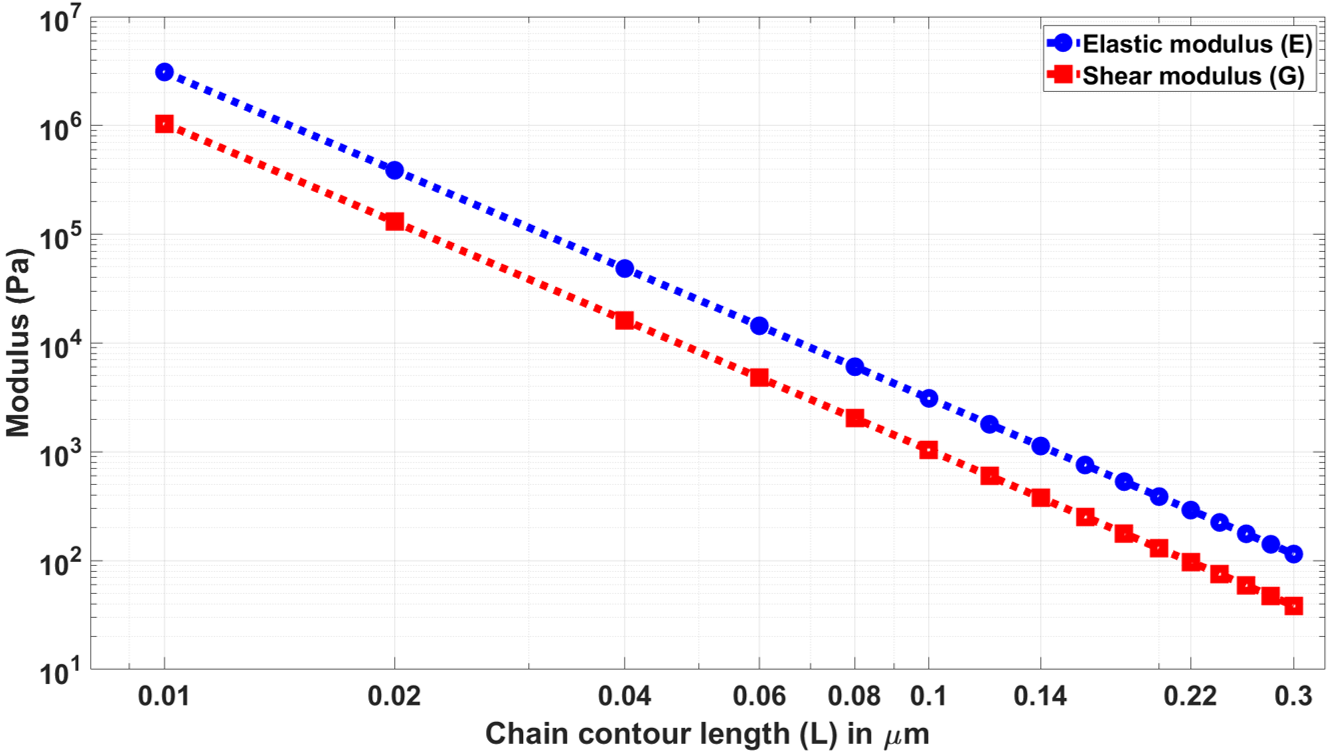}
    \caption{Elastic and shear moduli as a function of chain contour length; as expected from polymer theory, these scale as $a^{-3}$ \cite{rubinstein2003polymer}.}
    \label{fig:modulus_vs_chain_length}
\end{figure*}

Figure \ref{fig:modulus_vs_chain_length} shows the effect of chain contour length on the elastic moduli of the polymer network. 
The chain contour length $L$ is varied from $0.01 \unit{\micro\meter}$ to $0.3 \unit{\micro\meter}$ while keeping $N$ fixed.
We observe that both the elastic modulus $E$ and the shear modulus $G$ decrease with increased chain contour length.
An increase in chain contour length corresponds to an increase in the average molecular weight ($M_c$) between the cross-links, and these results are consistent with experiments that show that an increase in $M_c$ corresponds to a decrease in the elastic moduli \cite{lovell2003effect,wisotzki2014tailoring}. 
The range of elastic and shear moduli obtained using the model by varying chain contour length is consistent with the experimental values for polymer network soft matter such as elastomers and polymeric gels \cite{ zhao2023self, krajina2018active, li2014hybrid, li2020effect, darnell2013performance, nepal2023hierarchically,  
norman2021measuring, feig2018mechanically, ohm2021electrically, xu2023concurrent, coyle2018bio, li2014stiff, sun2022rheology, cai2011periodic, shan2015rigidity}.

\subsection{Interactions between Deformation and an Excluded Volume Instability}

We next examine an instability driven by the increase in the excluded volume parameter.
For computational feasability -- because we aim to numerically confirm that the instability is sharp by a large number of calculations near the instability -- we focus on 2-d systems; however, a few representative calculations suggest that 3-d is qualitatively similar.
Because it is in 2-d, the undeformed RVE is a square with 4 chains.

\begin{figure*}[ht!]
    {\includegraphics[width=0.94\textwidth]{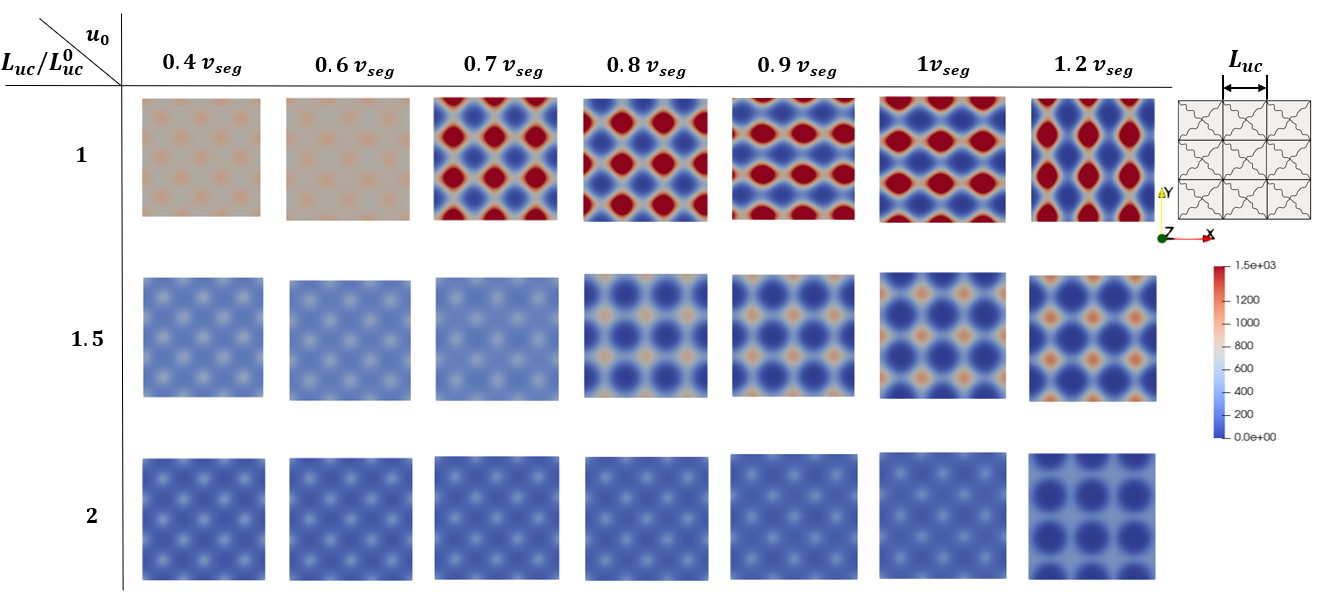}}
	\caption{
            Excluded volume-driven instability observed in 2-d, and the effect of equi-biaxial deformation. 
            Each subfigure shows the average segment density plotted over 9 RVEs, with each subfigure corresponding to different values of biaxial stretch $L_{uc}/L_{uc}^0$and $u_0$.
            The instability corresponds to a sharp transition in the chain configuration: it goes from being fairly uniform away from the crosslinking point to being concentrated along the horizontal and vertical directions.}
            As the stretch increases, the critical value of the excluded volume parameter at which the instability occurs also increases.
	\label{fig:pattern_instability}
\end{figure*}

Figure \ref{fig:pattern_instability} presents the average segment density field for various equi-biaxial stretches and various excluded volume parameter values.
For a fixed RVE stretch of $L_{uc}/L_{uc}^0 = 1$, where $L^0_{uc}=\sqrt2 \ aN^{1/2}$ in 2-d, we observe an instability at $u_0 \approx 0.7 \ v_{seg}$ $(v_{seg}=a^2$ in 2-d), leading to the localization of chains.
Physically, the chains strongly repel each other and hence are highly restricted in the volume available.
The instability is symmetry breaking, in that the originally square-symmetric chain configuration transitions to localize either vertically or horizontally from the original square symmetry; in our numerical simulations, we find that these occur essentially randomly due to numerical noise.
As noted above, the instability is a sharp transition.

We examine the effect of an imposed equi-biaxial stretch by setting $L_{uc}/L_{uc}^0$ to $1.5$ and $2$ respectively.
We notice that the critical values of $u_0$ for the instability are, respectively, $u_0 \approx 0.8 \ v_{seg}$ and $u_0 \approx 1.2 \ v_{seg}$.
This coupling between the deformation and chain localization suggests new routes to obtain patterning in polymer networks.

\section{Discussion}\label{sec: conclusion}
    
We have used the statistical field theory of polymers in combination with the 8-chain network averaging approach to study the mechanical response of polymer networks.
The framework of polymer field theory provides a physics-based approach to accounting for excluded volume interactions, which are imposed phenomenologically in micromechanical models.
In the absence of excluded volume interactions, we find that that the closed-form orientationally-averaged elastic response matches with the classical rubber elasticity \cite{treloar1975physics}.  
With excluded volume effects, self-consistent numerical solutions using finite elements find that the predicted elastic moduli are in line with typical polymer network gels; particularly, the linearized Poisson's ratio $\nu \simeq 0.4943$, which is very close to the incompressible limit $\nu\to 0.5$, without a phenomenological imposition of incompressibility.
Though the equilibrium state depends on the value of $u_0$, the incompressible behavior is independent of the specific value of $u_0$ for the values studied here. 
This can be physically understood by considering that $\nu$ is computed around the equilibrium stress-free state. 
Due to entropic effects, the chains tend to reduce their end-to-end distance and would collapse to a point, while the excluded volume effects prevent the chains from collapsing completely.
The equilibrium state is achieved when these opposing effects balance out. 
Around this equilibrium state, we find $\nu \simeq 0.5$, which reflects the role of the solvent in preventing further reduction in volume.
Despite the seeming presence of voids or open space in the density fields, the chains are constrained due to the excluded volume interactions; the voids will be occupied by the solvent which makes the system close to incompressible since the solvent cannot leave the polymer network upon deformation.
This is consistent with the results of \cite{pritchard2013swelling}, wherein it was found that $\nu\simeq 0.5$ at short times when the solvent has not had time to diffuse out of the polymer network; at longer times, the solvent diffuses out and the long-time equilibrium value of $\nu$ depends on the shear modulus.
While we have not considered this time-dependent behavior here, it is related to similar effects in poromechanics, namely the Terzaghi and Mandel model problems, wherein the stress response is closely tied to the drainage of the pore fluid \cite{biot1973nonlinear,coussy2004poromechanics,rudnicki2001coupled,karimi2022energetic,wang2000theory}.

Another interesting finding is that, despite the harmonic Gaussian chain as a starting point, there is an emergent strain-softening and strain-stiffening response that is characteristic of real polymer network gels, driven by the interplay between the excluded volume interactions and the entropy; it does not require chains with limiting extensibility -- such as the inverse Langevin approximation -- to model this behavior.
We also we find the emergence of a deformation-sensitive localization instability at large values of the excluded volume parameter.

A natural question for future is to examine the interplay between chain-scale instabilities such as microbuckling \cite{grekas2021cells,lakes1993microbuckling,sun2020fibrous} and the network-scale instabilities observed here.
We highlight, however, that these examples of instabilities are in fibrous networks, and it is possible that such instabilities occur here because of the regular network structure that has been assumed and will not appear in random networks.
An important challenge, however, is that the isolated harmonic Gaussian chain does not display buckling or other instabilities; other nonlinear chain models are required to capture this behavior.
Further, as noted in \cite{grasinger2021nonlinear}, electrical field interactions provide an effective compressive stiffness, and can induce new types of instabilities \cite{grasinger2021flexoelectricity}.
Incorporating
 chain models that go beyond the Gaussian approximation in polymer field theory is an interesting theoretical question.
Along similar lines, while we capture strain stiffening and strain softening without the use of a chain model with limiting extensibility -- such as the inverse Langevin model -- it would be interesting to incorporate such models in polymer field theory to enable studying the interplay between entropic, excluded volume, and limited extensibility effects.

    The concept of chain topology or entanglement is an important aspect that is not taken into consideration in this study. 
    As highlighted in \cite{davidson2013nonaffine,davidson2016nonaffine}, these effects can play a significant role in the response of polymer networks. 
    The mean-field framework employed in this study is unable to account for such effects directly. 
    However, our inclusion of excluded volume effects provides some insight into the effects of entanglement. 
    Additionally, we have observed that even without accounting for entanglements, excluded volume effects give rise to many interesting physical characteristics that are relevant to the response of real polymer networks. 
    It should be noted that while entanglements are crucial for large deformation response, the linearized properties such as the Poisson's ratio are expected to be relatively unaltered.

\section*{Code Availability}

A version of the code developed for this work is available at  \\ \url{https://github.com/pkhandag/polymer-network.git}

\begin{acknowledgments}
    We thank Carlos Garcia Cervera for useful discussions; 
    NSF XSEDE for computing resources provided by the Pittsburgh Supercomputing Center;
    AFRL for hosting visits by Kaushik Dayal;
    NSF (DMREF 1921857, DMS 2108784), ONR (N00014-18-1-2528), BSF (2018183), and AFOSR (MURI FA9550-18-1-0095) for financial support;
    and the anonymous reviewers for comments that improved the paper significantly.
\end{acknowledgments}


\bibliographystyle{unsrt}

\begin{thebibliography}{100}

\bibitem{deng2014electrets}
Qian Deng, Liping Liu, and Pradeep Sharma.
\newblock Electrets in soft materials: Nonlinearity, size effects, and giant
  electromechanical coupling.
\newblock {\em Physical Review E}, 90(1):012603, 2014.

\bibitem{chen2021interplay}
Lingling Chen, Xu~Yang, Binglei Wang, Shengyou Yang, Kaushik Dayal, and Pradeep
  Sharma.
\newblock The interplay between symmetry-breaking and symmetry-preserving
  bifurcations in soft dielectric films and the emergence of giant
  electro-actuation.
\newblock {\em Extreme Mechanics Letters}, 43:101151, 2021.

\bibitem{ahmadpoor2015flexoelectricity}
Fatemeh Ahmadpoor and Pradeep Sharma.
\newblock Flexoelectricity in two-dimensional crystalline and biological
  membranes.
\newblock {\em Nanoscale}, 7(40):16555--16570, 2015.

\bibitem{kim2016foam}
Oleg~V Kim, Xiaojun Liang, Rustem~I Litvinov, John~W Weisel, Mark~S Alber, and
  Prashant~K Purohit.
\newblock Foam-like compression behavior of fibrin networks.
\newblock {\em Biomechanics and modeling in mechanobiology}, 15(1):213--228,
  2016.

\bibitem{brown2009multiscale}
Andr{\'e}~EX Brown, Rustem~I Litvinov, Dennis~E Discher, Prashant~K Purohit,
  and John~W Weisel.
\newblock Multiscale mechanics of fibrin polymer: gel stretching with protein
  unfolding and loss of water.
\newblock {\em science}, 325(5941):741--744, 2009.

\bibitem{su2012semiflexible}
Tianxiang Su and Prashant~K Purohit.
\newblock Semiflexible filament networks viewed as fluctuating beam-frames.
\newblock {\em Soft Matter}, 8(17):4664--4674, 2012.

\bibitem{grasinger2021architected}
Matthew Grasinger and Kaushik Dayal.
\newblock Architected elastomer networks for optimal electromechanical
  response.
\newblock {\em Journal of the Mechanics and Physics of Solids}, 146:104171,
  2021.

\bibitem{grasinger2021nonlinear}
Matthew Grasinger, Carmel Majidi, and Kaushik Dayal.
\newblock Nonlinear statistical mechanics drives intrinsic electrostriction and
  volumetric torque in polymer networks.
\newblock {\em Physical Review E}, 103(4):042504, 2021.

\bibitem{grasinger2020statistical}
Matthew Grasinger and Kaushik Dayal.
\newblock Statistical mechanical analysis of the electromechanical coupling in
  an electrically-responsive polymer chain.
\newblock {\em Soft Matter}, 16(27):6265--6284, 2020.

\bibitem{cohen2016electroelasticity}
Noy Cohen, Kaushik Dayal, and Gal deBotton.
\newblock Electroelasticity of polymer networks.
\newblock {\em Journal of the Mechanics and Physics of Solids}, 92:105--126,
  2016.

\bibitem{darbaniyan2019designing}
Faezeh Darbaniyan, Kaushik Dayal, Liping Liu, and Pradeep Sharma.
\newblock Designing soft pyroelectric and electrocaloric materials using
  electrets.
\newblock {\em Soft matter}, 15(2):262--277, 2019.

\bibitem{grasinger2021flexoelectricity}
Matthew Grasinger, Kosar Mozaffari, and Pradeep Sharma.
\newblock Flexoelectricity in soft elastomers and the molecular mechanisms
  underpinning the design and emergence of giant flexoelectricity.
\newblock {\em Proceedings of the National Academy of Sciences}, 118(21), 2021.

\bibitem{grasinger2022group}
Matthew Grasinger.
\newblock Group action markov chain monte carlo for accelerated sampling of
  energy landscapes with discrete symmetries and energy barriers.
\newblock {\em arXiv preprint arXiv:2205.00028}, 2022.

\bibitem{markvicka2018autonomously}
Eric~J Markvicka, Michael~D Bartlett, Xiaonan Huang, and Carmel Majidi.
\newblock An autonomously electrically self-healing liquid metal--elastomer
  composite for robust soft-matter robotics and electronics.
\newblock {\em Nature materials}, 17(7):618--624, 2018.

\bibitem{bartlett2017high}
Michael~D Bartlett, Navid Kazem, Matthew~J Powell-Palm, Xiaonan Huang, Wenhuan
  Sun, Jonathan~A Malen, and Carmel Majidi.
\newblock High thermal conductivity in soft elastomers with elongated liquid
  metal inclusions.
\newblock {\em Proceedings of the National Academy of Sciences},
  114(9):2143--2148, 2017.

\bibitem{kazem2017soft}
Navid Kazem, Tess Hellebrekers, and Carmel Majidi.
\newblock Soft multifunctional composites and emulsions with liquid metals.
\newblock {\em Advanced Materials}, 29(27):1605985, 2017.

\bibitem{majidi2019soft}
Carmel Majidi.
\newblock Soft-matter engineering for soft robotics.
\newblock {\em Advanced Materials Technologies}, 4(2):1800477, 2019.

\bibitem{zhao2021modeling}
Yongyi Zhao, Pratik Khandagale, and Carmel Majidi.
\newblock Modeling electromechanical coupling of liquid metal embedded
  elastomers while accounting stochasticity in 3d percolation.
\newblock {\em Extreme Mechanics Letters}, 48:101443, 2021.

\bibitem{bartlett2019self}
Michael~D Bartlett, Michael~D Dickey, and Carmel Majidi.
\newblock Self-healing materials for soft-matter machines and electronics.
\newblock {\em NPG Asia Materials}, 11(1):1--4, 2019.

\bibitem{ford2019multifunctional}
Michael~J Ford, Cedric~P Ambulo, Teresa~A Kent, Eric~J Markvicka, Chengfeng
  Pan, Jonathan Malen, Taylor~H Ware, and Carmel Majidi.
\newblock A multifunctional shape-morphing elastomer with liquid metal
  inclusions.
\newblock {\em Proceedings of the National Academy of Sciences},
  116(43):21438--21444, 2019.

\bibitem{zolfaghari2020network}
Navid Zolfaghari, Pratik Khandagale, Michael~J Ford, Kaushik Dayal, and Carmel
  Majidi.
\newblock Network topologies dictate electromechanical coupling in liquid
  metal--elastomer composites.
\newblock {\em Soft Matter}, 16(38):8818--8825, 2020.

\bibitem{ohm2021electrically}
Yunsik Ohm, Chengfeng Pan, Michael~J Ford, Xiaonan Huang, Jiahe Liao, and
  Carmel Majidi.
\newblock An electrically conductive silver--polyacrylamide--alginate hydrogel
  composite for soft electronics.
\newblock {\em Nature Electronics}, 4(3):185--192, 2021.

\bibitem{malakooti2020liquid}
Mohammad~H Malakooti, Michael~R Bockstaller, Krzysztof Matyjaszewski, and
  Carmel Majidi.
\newblock Liquid metal nanocomposites.
\newblock {\em Nanoscale Advances}, 2(7):2668--2677, 2020.

\bibitem{ware2016localized}
Taylor~H Ware, John~S Biggins, Andreas~F Shick, Mark Warner, and Timothy~J
  White.
\newblock Localized soft elasticity in liquid crystal elastomers.
\newblock {\em Nature communications}, 7(1):1--7, 2016.

\bibitem{white2015programmable}
Timothy~J White and Dirk~J Broer.
\newblock Programmable and adaptive mechanics with liquid crystal polymer
  networks and elastomers.
\newblock {\em Nature materials}, 14(11):1087--1098, 2015.

\bibitem{ware2015voxelated}
Taylor~H Ware, Michael~E McConney, Jeong~Jae Wie, Vincent~P Tondiglia, and
  Timothy~J White.
\newblock Voxelated liquid crystal elastomers.
\newblock {\em Science}, 347(6225):982--984, 2015.

\bibitem{ambulo2017four}
Cedric~P Ambulo, Julia~J Burroughs, Jennifer~M Boothby, Hyun Kim, M~Ravi
  Shankar, and Taylor~H Ware.
\newblock Four-dimensional printing of liquid crystal elastomers.
\newblock {\em ACS applied materials \& interfaces}, 9(42):37332--37339, 2017.

\bibitem{mu2019sheath}
Jiuke Mu, M{\^o}nica~Jung De~Andrade, Shaoli Fang, Xuemin Wang, Enlai Gao,
  Na~Li, Shi~Hyeong Kim, Hongzhi Wang, Chengyi Hou, Qinghong Zhang, et~al.
\newblock Sheath-run artificial muscles.
\newblock {\em Science}, 365(6449):150--155, 2019.

\bibitem{saed2019molecularly}
Mohand~O Saed, Cedric~P Ambulo, Hyun Kim, Rohit De, Vyom Raval, Kyle Searles,
  Danyal~A Siddiqui, John Michael~O Cue, Mihaela~C Stefan, M~Ravi Shankar,
  et~al.
\newblock Molecularly-engineered, 4d-printed liquid crystal elastomer
  actuators.
\newblock {\em Advanced Functional Materials}, 29(3):1806412, 2019.

\bibitem{wie2016photomotility}
Jeong~Jae Wie, M~Ravi Shankar, and Timothy~J White.
\newblock Photomotility of polymers.
\newblock {\em Nature communications}, 7(1):1--8, 2016.

\bibitem{babaei2017steering}
Mahnoush Babaei, J~Arul Clement, Kaushik Dayal, and M~Ravi Shankar.
\newblock Steering with light: indexable photomotility in liquid crystalline
  polymers.
\newblock {\em RSC advances}, 7(83):52510--52516, 2017.

\bibitem{babaei2021torque}
Mahnoush Babaei, Junfeng Gao, Arul Clement, Kaushik Dayal, and M~Ravi Shankar.
\newblock Torque-dense photomechanical actuation.
\newblock {\em Soft Matter}, 17(5):1258--1266, 2021.

\bibitem{mooney1940theory}
Melvin Mooney.
\newblock A theory of large elastic deformation.
\newblock {\em Journal of applied physics}, 11(9):582--592, 1940.

\bibitem{ogden1997non}
Raymond~W Ogden.
\newblock {\em Non-linear elastic deformations}.
\newblock Courier Corporation, 1997.

\bibitem{gent1996new}
Alan~N Gent.
\newblock A new constitutive relation for rubber.
\newblock {\em Rubber chemistry and technology}, 69(1):59--61, 1996.

\bibitem{james1943theory}
Hubert~M James and Eugene Guth.
\newblock Theory of the elastic properties of rubber.
\newblock {\em The Journal of Chemical Physics}, 11(10):455--481, 1943.

\bibitem{flory1943statistical}
Paul~J Flory and John Rehner~Jr.
\newblock Statistical mechanics of cross-linked polymer networks i. rubberlike
  elasticity.
\newblock {\em The journal of chemical physics}, 11(11):512--520, 1943.

\bibitem{treloar1946elasticity}
LRG Treloar.
\newblock The elasticity of a network of long-chain molecules.—iii.
\newblock {\em Transactions of the Faraday Society}, 42:83--94, 1946.

\bibitem{treloar1954photoelastic}
LRG Treloar.
\newblock The photoelastic properties of short-chain molecular networks.
\newblock {\em Transactions of the Faraday Society}, 50:881--896, 1954.

\bibitem{arruda1993three}
Ellen~M Arruda and Mary~C Boyce.
\newblock A three-dimensional constitutive model for the large stretch behavior
  of rubber elastic materials.
\newblock {\em Journal of the Mechanics and Physics of Solids}, 41(2):389--412,
  1993.

\bibitem{wu1993improved}
PD~Wu and Erik Van Der~Giessen.
\newblock On improved network models for rubber elasticity and their
  applications to orientation hardening in glassy polymers.
\newblock {\em Journal of the Mechanics and Physics of Solids}, 41(3):427--456,
  1993.

\bibitem{miehe2004micro}
C~Miehe, Serdar G{\"o}ktepe, and F~Lulei.
\newblock A micro-macro approach to rubber-like materials—part i: the
  non-affine micro-sphere model of rubber elasticity.
\newblock {\em Journal of the Mechanics and Physics of Solids},
  52(11):2617--2660, 2004.

\bibitem{grasinger2023networks}
Matthew Grasinger.
\newblock Polymer networks which locally rotate to accommodate stresses,
  torques, and deformation.
\newblock {\em Journal of the Mechanics and Physics of Solids}, 175:105289,
  2023.

\bibitem{ronca1975approach}
G~Ronca and G~Allegra.
\newblock An approach to rubber elasticity with internal constraints.
\newblock {\em The Journal of Chemical Physics}, 63(11):4990--4997, 1975.

\bibitem{flory1976statistical}
Paul~J Flory.
\newblock Statistical thermodynamics of random networks.
\newblock {\em Proceedings of the Royal Society of London. A. Mathematical and
  Physical Sciences}, 351(1666):351--380, 1976.

\bibitem{flory1977theory}
PJ~Flory.
\newblock Theory of elasticity of polymer networks. the effect of local
  constraints on junctions.
\newblock {\em The Journal of Chemical Physics}, 66(12):5720--5729, 1977.

\bibitem{erman1978theory}
Burak Erman and Paul~J Flory.
\newblock Theory of elasticity of polymer networks. ii. the effect of geometric
  constraints on junctions.
\newblock {\em The Journal of Chemical Physics}, 68(12):5363--5369, 1978.

\bibitem{flory1982theory}
Paul~J Flory and Burak Erman.
\newblock Theory of elasticity of polymer networks. 3.
\newblock {\em Macromolecules}, 15(3):800--806, 1982.

\bibitem{deam1976theory}
RT~Deam and Samuel~Frederick Edwards.
\newblock The theory of rubber elasticity.
\newblock {\em Philosophical Transactions of the Royal Society of London.
  Series A, Mathematical and Physical Sciences}, 280(1296):317--353, 1976.

\bibitem{edwards1988tube}
SF~Edwards and Th~A Vilgis.
\newblock The tube model theory of rubber elasticity.
\newblock {\em Reports on Progress in Physics}, 51(2):243, 1988.

\bibitem{heinrich1983strength}
G~Heinrich and E~Straube.
\newblock On the strength and deformation dependence of the tube-like
  topological constraints of polymer networks, melts and concentrated
  solutions. i. the polymer network case.
\newblock {\em Acta polymerica}, 34(9):589--594, 1983.

\bibitem{heinrich1984strength}
G~Heinrich and E~Straube.
\newblock On the strength and deformation dependence of the tube-like
  topological constraints of polymer networks, melts and concentrated
  solutions. ii. polymer melts and concentrated solutions.
\newblock {\em Acta polymerica}, 35(2):115--119, 1984.

\bibitem{heinrich1988rubber}
G~Heinrich, E~Straube, and G~Helmis.
\newblock Rubber elasticity of polymer networks: Theories.
\newblock {\em Polymer physics}, pages 33--87, 1988.

\bibitem{de1969some}
P-G De~Gennes.
\newblock Some conformation problems for long macromolecules.
\newblock {\em Reports on Progress in Physics}, 32(1):187, 1969.

\bibitem{de1979scaling}
Pierre-Gilles De~Gennes.
\newblock {\em Scaling concepts in polymer physics}.
\newblock Cornell university press, 1979.

\bibitem{matsen2006self}
Mark~W Matsen.
\newblock Self-consistent field theory and its applications.
\newblock {\em Soft Matter}, 1, 2006.

\bibitem{doi1986theory}
Masao Doi and Samuel~Frederick Edwards.
\newblock {\em The theory of polymer dynamics}.
\newblock oxford university press, 1986.

\bibitem{fredrickson2006equilibrium}
Glenn Fredrickson.
\newblock {\em The equilibrium theory of inhomogeneous polymers}.
\newblock Oxford University Press, 2006.

\bibitem{fredrickson2002field}
Glenn~H Fredrickson, Venkat Ganesan, and Fran{\c{c}}ois Drolet.
\newblock Field-theoretic computer simulation methods for polymers and complex
  fluids.
\newblock {\em Macromolecules}, 35(1):16--39, 2002.

\bibitem{chantawansri2007self}
Tanya~L Chantawansri, August~W Bosse, Alexander Hexemer, Hector~D Ceniceros,
  Carlos~J Garcia-Cervera, Edward~J Kramer, and Glenn~H Fredrickson.
\newblock Self-consistent field theory simulations of block copolymer assembly
  on a sphere.
\newblock {\em Physical Review E}, 75(3):031802, 2007.

\bibitem{lennon2008free}
Erin~M Lennon, Kirill Katsov, and Glenn~H Fredrickson.
\newblock Free energy evaluation in field-theoretic polymer simulations.
\newblock {\em Physical review letters}, 101(13):138302, 2008.

\bibitem{drolet1999combinatorial}
Fran{\c{c}}ois Drolet and Glenn~H Fredrickson.
\newblock Combinatorial screening of complex block copolymer assembly with
  self-consistent field theory.
\newblock {\em Physical Review Letters}, 83(21):4317, 1999.

\bibitem{sides2006hybrid}
Scott~W Sides, Bumjoon~J Kim, Edward~J Kramer, and Glenn~H Fredrickson.
\newblock Hybrid particle-field simulations of polymer nanocomposites.
\newblock {\em Physical review letters}, 96(25):250601, 2006.

\bibitem{ackerman2017finite}
David~M Ackerman, Kris Delaney, Glenn~H Fredrickson, and Baskar
  Ganapathysubramanian.
\newblock A finite element approach to self-consistent field theory
  calculations of multiblock polymers.
\newblock {\em Journal of Computational Physics}, 331:280--296, 2017.

\bibitem{fredrickson2007computational}
Glenn~H Fredrickson.
\newblock Computational field theory of polymers: opportunities and challenges.
\newblock {\em Soft Matter}, 3(11):1329--1334, 2007.

\bibitem{delaney2016recent}
Kris~T Delaney and Glenn~H Fredrickson.
\newblock Recent developments in fully fluctuating field-theoretic simulations
  of polymer melts and solutions.
\newblock {\em The Journal of Physical Chemistry B}, 120(31):7615--7634, 2016.

\bibitem{cochran2006stability}
Eric~W Cochran, Carlos~J Garcia-Cervera, and Glenn~H Fredrickson.
\newblock Stability of the gyroid phase in diblock copolymers at strong
  segregation.
\newblock {\em Macromolecules}, 39(7):2449--2451, 2006.

\bibitem{lennon2008numerical}
Erin~M Lennon, George~O Mohler, Hector~D Ceniceros, Carlos~J
  Garc{\'\i}a-Cervera, and Glenn~H Fredrickson.
\newblock Numerical solutions of the complex langevin equations in polymer
  field theory.
\newblock {\em Multiscale Modeling \& Simulation}, 6(4):1347--1370, 2008.

\bibitem{matsen1994stable}
Mark~W Matsen and Michael Schick.
\newblock Stable and unstable phases of a diblock copolymer melt.
\newblock {\em Physical Review Letters}, 72(16):2660, 1994.

\bibitem{matsen2001standard}
Mark~W Matsen.
\newblock The standard gaussian model for block copolymer melts.
\newblock {\em Journal of Physics: Condensed Matter}, 14(2):R21, 2001.

\bibitem{schmid2013self}
Friederike Schmid.
\newblock Self-consistent field approach for cross-linked copolymer materials.
\newblock {\em Physical Review Letters}, 111(2):028303, 2013.

\bibitem{treloar1975physics}
Leslie Ronald~George Treloar.
\newblock {\em The physics of rubber elasticity}.
\newblock Oxford University Press, USA, 1975.

\bibitem{davidson2013nonaffine}
Jacob~D Davidson and Nakhiah~C Goulbourne.
\newblock A nonaffine network model for elastomers undergoing finite
  deformations.
\newblock {\em Journal of the Mechanics and Physics of Solids},
  61(8):1784--1797, 2013.

\bibitem{davidson2016nonaffine}
Jacob~D Davidson and NC~Goulbourne.
\newblock Nonaffine chain and primitive path deformation in crosslinked
  polymers.
\newblock {\em Modelling and Simulation in Materials Science and Engineering},
  24(6):065002, 2016.

\bibitem{rubinstein1997nonaffine}
Michael Rubinstein and Sergei Panyukov.
\newblock Nonaffine deformation and elasticity of polymer networks.
\newblock {\em Macromolecules}, 30(25):8036--8044, 1997.

\bibitem{rubinstein2002elasticity}
Michael Rubinstein and Sergei Panyukov.
\newblock Elasticity of polymer networks.
\newblock {\em Macromolecules}, 35(17):6670--6686, 2002.

\bibitem{zimm1953excluded}
BeHe Zimm, WH~Stockmayer, and M~Fixman.
\newblock Excluded volume in polymer chains.
\newblock {\em The Journal of Chemical Physics}, 21(10):1716--1723, 1953.

\bibitem{langtangen2017solving}
Hans~Petter Langtangen and Anders Logg.
\newblock {\em Solving PDEs in python: the FEniCS tutorial I}.
\newblock Springer Nature, 2017.

\bibitem{milner2009chi}
Scott~T Milner, Martin-Daniel Lacasse, and William~W Graessley.
\newblock Why $\chi$ is seldom zero for polymer- solvent mixtures.
\newblock {\em Macromolecules}, 42(3):876--886, 2009.

\bibitem{nistane2022estimation}
Janhavi Nistane, Lihua Chen, Youngjoo Lee, Ryan Lively, and Rampi Ramprasad.
\newblock Estimation of the flory-huggins interaction parameter of
  polymer-solvent mixtures using machine learning.
\newblock {\em MRS Communications}, pages 1--7, 2022.

\bibitem{wittmer2007intramolecular}
JP~Wittmer, Philippe Beckrich, Hendrik Meyer, Anna Cavallo, Albert Johner, and
  Joerg Baschnagel.
\newblock Intramolecular long-range correlations in polymer melts: The
  segmental size distribution and its moments.
\newblock {\em Physical Review E}, 76(1):011803, 2007.

\bibitem{lang2015conformations}
Michael Lang, Michael Rubinstein, and Jens-Uwe Sommer.
\newblock Conformations of a long polymer in a melt of shorter chains:
  Generalizations of the flory theorem.
\newblock {\em ACS Macro Letters}, 4(2):177--181, 2015.

\bibitem{hong2008theory}
Wei Hong, Xuanhe Zhao, Jinxiong Zhou, and Zhigang Suo.
\newblock A theory of coupled diffusion and large deformation in polymeric
  gels.
\newblock {\em Journal of the Mechanics and Physics of Solids},
  56(5):1779--1793, 2008.

\bibitem{kamata2015design}
Hiroyuki Kamata, Xiang Li, Ung-il Chung, and Takamasa Sakai.
\newblock Design of hydrogels for biomedical applications.
\newblock {\em Advanced healthcare materials}, 4(16):2360--2374, 2015.

\bibitem{tanaka1979kinetics}
Toyoichi Tanaka and David~J Fillmore.
\newblock Kinetics of swelling of gels.
\newblock {\em The Journal of Chemical Physics}, 70(3):1214--1218, 1979.

\bibitem{jagur2010polymeric}
Joseph Jagur-Grodzinski.
\newblock Polymeric gels and hydrogels for biomedical and pharmaceutical
  applications.
\newblock {\em Polymers for Advanced Technologies}, 21(1):27--47, 2010.

\bibitem{norman2021measuring}
Michael~DA Norman, Silvia~A Ferreira, Geraldine~M Jowett, Laurent Bozec, and
  Eileen Gentleman.
\newblock Measuring the elastic modulus of soft culture surfaces and
  three-dimensional hydrogels using atomic force microscopy.
\newblock {\em Nature Protocols}, 16(5):2418--2449, 2021.

\bibitem{feig2018mechanically}
Vivian~R Feig, Helen Tran, Minah Lee, and Zhenan Bao.
\newblock Mechanically tunable conductive interpenetrating network hydrogels
  that mimic the elastic moduli of biological tissue.
\newblock {\em Nature communications}, 9(1):2740, 2018.

\bibitem{xu2023concurrent}
Shuai Xu, Zidi Zhou, Zishun Liu, and Pradeep Sharma.
\newblock Concurrent stiffening and softening in hydrogels under dehydration.
\newblock {\em Science Advances}, 9(1):eade3240, 2023.

\bibitem{coyle2018bio}
Stephen Coyle, Carmel Majidi, Philip LeDuc, and K~Jimmy Hsia.
\newblock Bio-inspired soft robotics: Material selection, actuation, and
  design.
\newblock {\em Extreme Mechanics Letters}, 22:51--59, 2018.

\bibitem{li2014stiff}
Jianyu Li, Zhigang Suo, and Joost~J Vlassak.
\newblock Stiff, strong, and tough hydrogels with good chemical stability.
\newblock {\em Journal of Materials Chemistry B}, 2(39):6708--6713, 2014.

\bibitem{sun2022rheology}
Chuanpeng Sun and Prashant~K Purohit.
\newblock Rheology of fibrous gels under compression.
\newblock {\em Extreme Mechanics Letters}, 54:101757, 2022.

\bibitem{cai2011periodic}
Shengqiang Cai, Derek Breid, Alfred~J Crosby, Zhigang Suo, and John~W
  Hutchinson.
\newblock Periodic patterns and energy states of buckled films on compliant
  substrates.
\newblock {\em Journal of the Mechanics and Physics of Solids},
  59(5):1094--1114, 2011.

\bibitem{shan2015rigidity}
Wanliang Shan, Stuart Diller, Abbas Tutcuoglu, and Carmel Majidi.
\newblock Rigidity-tuning conductive elastomer.
\newblock {\em Smart Materials and Structures}, 24(6):065001, 2015.

\bibitem{lovell2003effect}
Lale~G Lovell and Christopher~N Bowman.
\newblock The effect of kinetic chain length on the mechanical relaxation of
  crosslinked photopolymers.
\newblock {\em Polymer}, 44(1):39--47, 2003.

\bibitem{wisotzki2014tailoring}
Emilia~I Wisotzki, Marcel Hennes, Carsten Schuldt, Florian Engert, Wolfgang
  Knolle, Ulrich Decker, Josef~A K{\"a}s, Mareike Zink, and Stefan~G Mayr.
\newblock Tailoring the material properties of gelatin hydrogels by high energy
  electron irradiation.
\newblock {\em Journal of Materials Chemistry B}, 2(27):4297--4309, 2014.

\bibitem{zhao2023self}
Yongyi Zhao, Yunsik Ohm, Jiahe Liao, Yichi Luo, Huai-Yu Cheng, Phillip Won,
  Peter Roberts, Manuel~Reis Carneiro, Mohammad~F Islam, Jung~Hyun Ahn, et~al.
\newblock A self-healing electrically conductive organogel composite.
\newblock {\em Nature Electronics}, pages 1--10, 2023.

\bibitem{krajina2018active}
Brad~A Krajina, Audrey Zhu, Sarah~C Heilshorn, and Andrew~J Spakowitz.
\newblock Active dna olympic hydrogels driven by topoisomerase activity.
\newblock {\em Physical review letters}, 121(14):148001, 2018.

\bibitem{li2014hybrid}
Jianyu Li, Widusha~RK Illeperuma, Zhigang Suo, and Joost~J Vlassak.
\newblock Hybrid hydrogels with extremely high stiffness and toughness.
\newblock {\em ACS Macro Letters}, 3(6):520--523, 2014.

\bibitem{li2020effect}
Ziqian Li, Zishun Liu, Teng~Yong Ng, and Pradeep Sharma.
\newblock The effect of water content on the elastic modulus and fracture
  energy of hydrogel.
\newblock {\em Extreme Mechanics Letters}, 35:100617, 2020.

\bibitem{darnell2013performance}
Max~C Darnell, Jeong-Yun Sun, Manav Mehta, Christopher Johnson, Praveen~R
  Arany, Zhigang Suo, and David~J Mooney.
\newblock Performance and biocompatibility of extremely tough
  alginate/polyacrylamide hydrogels.
\newblock {\em Biomaterials}, 34(33):8042--8048, 2013.

\bibitem{nepal2023hierarchically}
Dhriti Nepal, Saewon Kang, Katarina~M Adstedt, Krishan Kanhaiya, Michael~R
  Bockstaller, L~Catherine Brinson, Markus~J Buehler, Peter~V Coveney, Kaushik
  Dayal, Jaafar~A El-Awady, et~al.
\newblock Hierarchically structured bioinspired nanocomposites.
\newblock {\em Nature materials}, 22(1):18--35, 2023.

\bibitem{rubinstein2003polymer}
Michael Rubinstein and Ralph~H Colby.
\newblock {\em Polymer physics}, volume~23.
\newblock Oxford university press, 2003.

\bibitem{pritchard2013swelling}
Robyn~H Pritchard and Eugene~M Terentjev.
\newblock Swelling and de-swelling of gels under external elastic deformation.
\newblock {\em Polymer}, 54(26):6954--6960, 2013.

\bibitem{biot1973nonlinear}
Maurice~A Biot.
\newblock Nonlinear and semilinear rheology of porous solids.
\newblock {\em Journal of Geophysical Research}, 78(23):4924--4937, 1973.

\bibitem{coussy2004poromechanics}
Olivier Coussy.
\newblock {\em Poromechanics}.
\newblock John Wiley \& Sons, 2004.

\bibitem{rudnicki2001coupled}
JW~Rudnicki.
\newblock Coupled deformation-diffusion effects in the mechanics of faulting
  and failure of geomaterials.
\newblock {\em Appl. Mech. Rev.}, 54(6):483--502, 2001.

\bibitem{karimi2022energetic}
Mina Karimi, Mehrdad Massoudi, Noel Walkington, Matteo Pozzi, and Kaushik
  Dayal.
\newblock Energetic formulation of large-deformation poroelasticity.
\newblock {\em International Journal for Numerical and Analytical Methods in
  Geomechanics}, 46(5):910--932, 2022.

\bibitem{wang2000theory}
Herbert Wang.
\newblock {\em Theory of linear poroelasticity with applications to
  geomechanics and hydrogeology}, volume~2.
\newblock Princeton university press, 2000.

\bibitem{grekas2021cells}
Georgios Grekas, Maria Proestaki, Phoebus Rosakis, Jacob Notbohm, Charalambos
  Makridakis, and Guruswami Ravichandran.
\newblock Cells exploit a phase transition to mechanically remodel the fibrous
  extracellular matrix.
\newblock {\em Journal of the Royal Society Interface}, 18(175):20200823, 2021.

\bibitem{lakes1993microbuckling}
R~Lakes, P~Rosakis, and A~Ruina.
\newblock Microbuckling instability in elastomeric cellular solids.
\newblock {\em Journal of materials science}, 28(17):4667--4672, 1993.

\bibitem{sun2020fibrous}
Chuanpeng Sun, Irina~N Chernysh, John~W Weisel, and Prashant~K Purohit.
\newblock Fibrous gels modelled as fluid-filled continua with double-well
  energy landscape.
\newblock {\em Proceedings of the Royal Society A}, 476(2244):20200643, 2020.

\end{thebibliography}

\end{document}